\documentclass{aa}
\usepackage{graphicx}
\def\kmps{km\,s$^{-1}$}
\def\ergcmsqs{erg\,cm$^{-2}$\,s$^{-1}$}
\def\fxo{$F_{\rm X}/F_{\rm opt}$}

\begin{document}
%\thesaurus {06()}

\title{The census of cataclysmic variables in 
the ROSAT Bright Survey\thanks{Based 
in part on observations performed at the Eurpean Southern Observatory, La Silla, 
Chile, under programmes 60.B-0107 and 66.A-0664}}
%\subtitle{The census of cataclysmic variables in the ROSAT Bright Survey}

\author{A.D. Schwope\inst{1}
\and
H. Brunner\inst{1,3}
\and
D. Buckley\inst{2}
\and
J. Greiner\inst{1,3}
\and
K. v.d. Heyden\inst{2,4}
\and
S. Neizvestny\inst{5}
\and
S. Potter\inst{2}
\and
R. Schwarz\inst{1,6}\thanks{Visiting astronomer,
              German-Spanish Astro\-no\-mi\-cal Center, Calar Alto, 
              operated by the Max-Planck-Institut f\"ur Astronomie, 
              Heidelberg, jointly with the Spanish National Commission 
              for Astronomy.}
}
\institute{Astrophysikalisches Institut Potsdam, An der Sternwarte 16, 14482 Potsdam, Germany
\and
South African Astronomical Observatory, PO Box 9, Observatory 7935, Cape 
     Town, RSA
\and
Max-Planck Institut f\"{u}r extraterrestrische Physik, Giessenbachstrasse, 85740 Garching, Germany
\and
SRON, Sorbonnelaan 2, 3584 CA Utrecht, The Netherlands
\and
SAO RAS, Nizhnij Arkhyz, Zelenchukskaya,
Karachaevo-Cherkesia, Russia, 357147
\and
Universit\"ats-Sternwarte G\"ottingen, Geismarlandstr.~11, 
37083 G\"ottingen, Germany
}
\date{Received; accepted } 

\abstract{We give an identification summary and results of 
polarimetric, photometric and spectroscopic follow-up
observations of new, X-ray bright 
cataclysmic variables. These were identified as optical counterparts of
high galactic latitude sources in the ROSAT All-Sky Survey. This 
optical identification programme is termed the ROSAT Bright Survey (RBS)
and represents the first complete soft X-ray selected, flux-limited sample
of CVs at high galactic latitude (survey area $\sim 20400$\,sq.deg.).
The systems described here escaped previous identification programmes 
since these surveys were designed to identify even brighter than 
ours or particularly
soft X-ray sources. Among the 11 new RBS-CVs we find 6 magnetic systems
of AM Herculis type, 4 dwarf novae (among them one candidate), and one 
particularly bright system of 
uncertain nature, tentatively identified as dwarf nova or symbiotic binary.
Orbital periods could be determined for all magnetic systems which range
from 87.1\,min to 187.7\,min. Three of the new dwarf novae have moderate 
to high inclination and two of them might be eclipsing. Using non-magnetic
systems only we derive a space density of CVs of 
$\sim 3\times 10^{-5}$\,pc$^{-3}$. This limit rests on the two new
nearby, low-luminosity systems RBS0490 and
RBS1955, with estimated distances
of 30\,pc only and luminosities below $10^{30}$\,erg\,s$^{-1}$.
\keywords{Surveys -- X-rays: binaries --  stars: cataclysmic variables}}
\maketitle

\section{Introduction}
In the ROSAT Bright Survey (RBS) we have completely optically 
identified all high-galactic latitude sources detected in the 
ROSAT All-Sky Survey with mean count rate above 0.2\,s$^{-1}$ 
(Schwope et al.~2000). About one third of all sources are of 
galactic origin. Most of them are stars with active coronae. 
Minority populations among the galactic constituents are
interacting binaries of different flavour (neutron star accretors,
white dwarf accretors, magnetic and non-magnetic systems, 
super-soft sources SSSs). With a survey area of more than
$20000$\,square degrees and its flux limit of about
$2.5\times 10^{-12}$\,\ergcmsqs\ 
the RBS represents the deepest complete,
soft X-ray selected, flux-limited sample of CVs 
(and other types of X-ray emitters)
at high galactic latitude, important for population
statistics and evolution scenarios of various kinds
of X-ray emitters  (see e.g.~Miyaji et al.~2001 for the 
X-ray luminosity function of AGNs based on the RBS and
other X-ray surveys).

\begin{table*}[th]
\caption{X-ray and optical summary information about new RBS-CVs. Listed are the 
RBS catalog number, the RASS catalog number containing the X-ray coordinates, the RASS
count rate and hardness ratio HR1, the coordinates of the optical counterpart, the 
magnitude in the optical range, the orbital period (if known), the X-ray to optical flux ratio 
and the suggested sub-class of interacting binary. The typical magnitude uncertainty 
is 0.1--0.2 mag.\label{t:list}}
\begin{tabular}{rllr@{$\pm$}l|rrccc}
\hline
RBS-No & 1RXS J & CR & \multicolumn{2}{c|}{HR1}& RA(2000) & DEC(2000) & $m_V$ & $P_{\rm orb}$ & Type\\
 & &[s$^{-1}$] & & & & & [mag] & [min] &  \\
\hline
206  &012851.9$-$233931  & 0.34 & $-0.84$ & 0.04 &01:28:52.3 & --23:39:44 &$17\fm5-18\fm0^{(a)}/ 18\fm9^{(b)}$&92.8& AM\\
324  &023052.9$-$684203  & 0.26 & $-0.70$ & 0.09 &02:30:50.9 & --68:42:06 &$16.0 - 18.0^{(a)}$&181.8 & AM\\
372  &025538.2$-$224655  & 0.23 & $-0.11$ & 0.13 &02:55:37.9 & --22:47:02 &$17\fm8$& --  &  DN\\
490  &035410.4$-$165244  & 0.23 & $+0.57$ & 0.08 &03:54:10.3 & --16:52:45 &$16\fm0$& --  &  DN\\
541  &042555.8$-$194534  & 0.23 & $+0.18$ & 0.12 &04:25:55.2 & --19:45:30 &$16\fm9$& 87.1&  AM\\ 
696  &082051.2$+$493433  & 0.25 & $-0.93$ & 0.03 &08:20:50.9 &  +49:34:31 &$18\fm0 - 19\fm3^{(a)}$& 99.4  & AM\\
1411 &143703.5$+$234236  & 0.25 & $+0.57$ & 0.08 &14:37:03.4 &  +23:42:27 &$19\fm1$& --  &  DN\\ 
1563 &161008.0$+$035222  & 0.54 & $-0.41$ & 0.11 &16:10:07.5 &  +03:52:33 &$15\fm9$&187.7&  AM\\
1735 &211540.9$-$584045  & 0.38 & $-0.02$ & 0.10 &21:15:40.6 & --58:40:52 &$17\fm8$     &110.8&  AM\\
1955 &230949.6$+$213523  & 0.21 & $+0.20$ & 0.10 &23:09:49.2 &  +21:35:19 &$15\fm6$& --  &  DN/Symb.\\
1969 &231532.3$-$304855  & 0.28 & $+0.31$ & 0.22 &23:15:31.7 & --30:48:48 &$16\fm6$& $84.2^{(c)}$& DN\\ 
\hline
%378  &02:59:31.2 & --04:10:52 & 0.61 &   & $+0.07$ & 0.16 \\ 
\hline
\end{tabular}
\\
$^{(a)}$ orbital high state variability \\
$^{(b)}$ low state magnitude\\
$^{(c)}$ Chen et al.~(2001)\\
\end{table*}

Due to the soft spectral band of the PSPC detector, 
ROSAT identifications revealed 
a large number of new magnetic CVs, most of AM Herculis subtype (polars),
and the SSSs as a new class of X-ray emitters. 
Polars are strongly 
magnetic CV where the magnetic field keeps both stars
(in most cases) in synchronous rotation, inhibits the formation 
of an accretion disk and leads to intensive 
cyclotron radiation in the optical and X-ray radiation 
from the accretion hot spot on the white dwarf.
Compared to pre-ROSAT times with less than 20 systems, the number 
of polars has increased to almost 70, most of them identified as counterparts
of very soft RASS or WFC-sources (Pye et al.~1995, Thomas et al.~1998,
Beuermann et al.~1999). The discovery of new systems and the 
determination of their basic parameters (orbital period, masses) 
is of importance for the evolution of this kind of binary. 
The number density
of the CVs in general is still highly uncertain. Their observed
minimum period is far from being compatible with current theoretical 
models and the presence or absence of the 2--3 hour period gap
and the likely causes for its presence or absence are a matter of debate.

However the restriction to soft X-ray sources {\it a priori} excludes 
the vast majority of CVs to be detected: normal disk systems in quiescence 
which have intrinsically hard spectra. Population synthesis 
using standard CV evolution predicts a space density of 
$\sim 10^{-5}$~pc$^{-3}$ (de Kool 1992) to 
$\sim 10^{-4}$~pc$^{-3}$ (Politano 1996), with the consequence 
that about 10 to 100 undetected CVs should be hiding in the local vicinity (50\,pc) 
of the Sun. Altough most of these sources have  been evolved to 
short periods, and accrete only at low $\dot{M}$, they should still emit 
$L_{\rm X} \sim 10^{30}$~erg~s$^{-1}$ and therefore show up in any medium deep X-ray  
survey.
The only pre-ROSAT medium deep X-ray search for CVs, 
the {\it Einstein} galactic plane survey (Hertz et al.~1990) detected
3 faint disk CVs and and one AM Herculis system in a 144 sqr degree field. 
The implied space density was $(2-3) \times 10^{-5}$~pc$^{-3}$ in agreement with the
theoretical predictions, but a factor of 5 to 10 higher than estimates 
based other detection criteria (variability, spectra, color) (see e.g Patterson 
1984).

DQ Herculis stars or intermediate polars are another class of CVs 
which are X-ray bright.  
They posses significant magnetic fields but the magnetic moment 
is too weak to synchronize both stars. Mass transfer may include 
an accretion disk and the systems have in general hard X-ray spectra.
Most DQ 
Herculis stars have periods above 3 hours and it's not clear whether
they later evolve into AM Herculis stars or form a largely unrecognized 
population. 
Thus an unbiased sample of CVs is crucial to validate our current 
understanding of their evolution.

%The likely identification of the more than 2000 individual RBS-sources 
%was given in the catalogue paper by Schwope et al.~(2000), the 
%observational data on which the classification was based
%was not presented there. 

For several of the new CVs found in the RBS (Schwope et al.~2000)
follow-up observations were 
initiated. In this paper we present the current status of the 
observations of these systems, including the results of photometric,
polarimetric and spectroscopic observations.
Results of our initial studies of RBS0206 and RBS1735, now called
EQ Cet und CD Ind, respectively, were presented 
earlier (Schwope et al.~1997, 1999). The main properties 
of the objects under investigation in this paper are summarized 
in Table~\ref{t:list}.

The paper is organized in the following way. We briefly describe our 
optical observations and then present the results and implications
in sections devoted to individual objects, separating magnetic 
and non-magnetic systems. The paper is closed with 
a short discussion of all CVs in the RBS.

\section{Observational summary of new interacting binaries found in the RBS}
%\section{Optical observations}
The spectra shown in this paper were obtained 
as part of the identification program of all RBS-sources (Schwope et al.~2000)
using the 6m-telescope at Zelenchukskaya (equipped with the 
UAGS+CCD prime focus spectrograph), the ESO 3.6m-telescope equipped
with EFOSC2, the Calar Alto 3.5m telescope with MOSCA,
and the ESO 1.5m-telescope equipped with a Boller \& Chivens
spectrograph. Low-resolution grisms or gratings were used in all cases, 
which gave spectra covering the whole optical wavelength range 
with moderate to low resolution (8 -- 20\,\AA\ FWHM). Follow-up spectroscopy,
partly with higher resolution (4\,\AA\ FWHM)
was obtained with the SAAO 1.9m telescope on
several occasions in 1998 and 1999.

Follow-up photometry 
and polarimetry was performed with the 
ESO/Dutch 90cm telescope with CCD camera, the AIP 70cm telescope with CCD
camera, and the SAAO 1.9m telescope with the UCT polarimeter.
More detailed information about the observations is given in the
sections on the individual targets.

Finding charts of the new CVs are reproduced in the Appendix. All charts
have a size of $5 \times 5$ arcmin, North at the top and East to the left.
The CVs are marked by short dashes.

\begin{figure}
\resizebox{\hsize}{!}
{\includegraphics[bbllx=50pt,bblly=77pt,bburx=530pt,bbury=774pt,angle=-90,clip=]{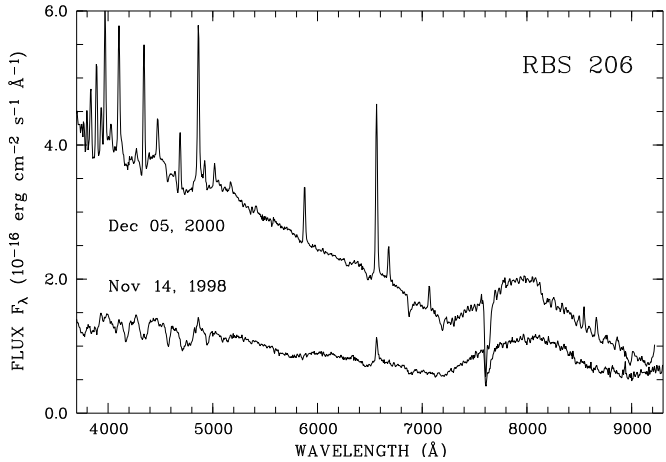}}
\caption{Single low-state discovery spectrum and mean 
high-state spectrum of RBS0206 taken during follow-up observations 
with the ESO 3.6m telescope. Observation dates are indicated in the figure.\label{f:s206}
}
\end{figure}

\begin{figure}[th]
\resizebox{\hsize}{!}
{\includegraphics[bbllx=52pt,bblly=179pt,bburx=333pt,bbury=767pt,clip=]{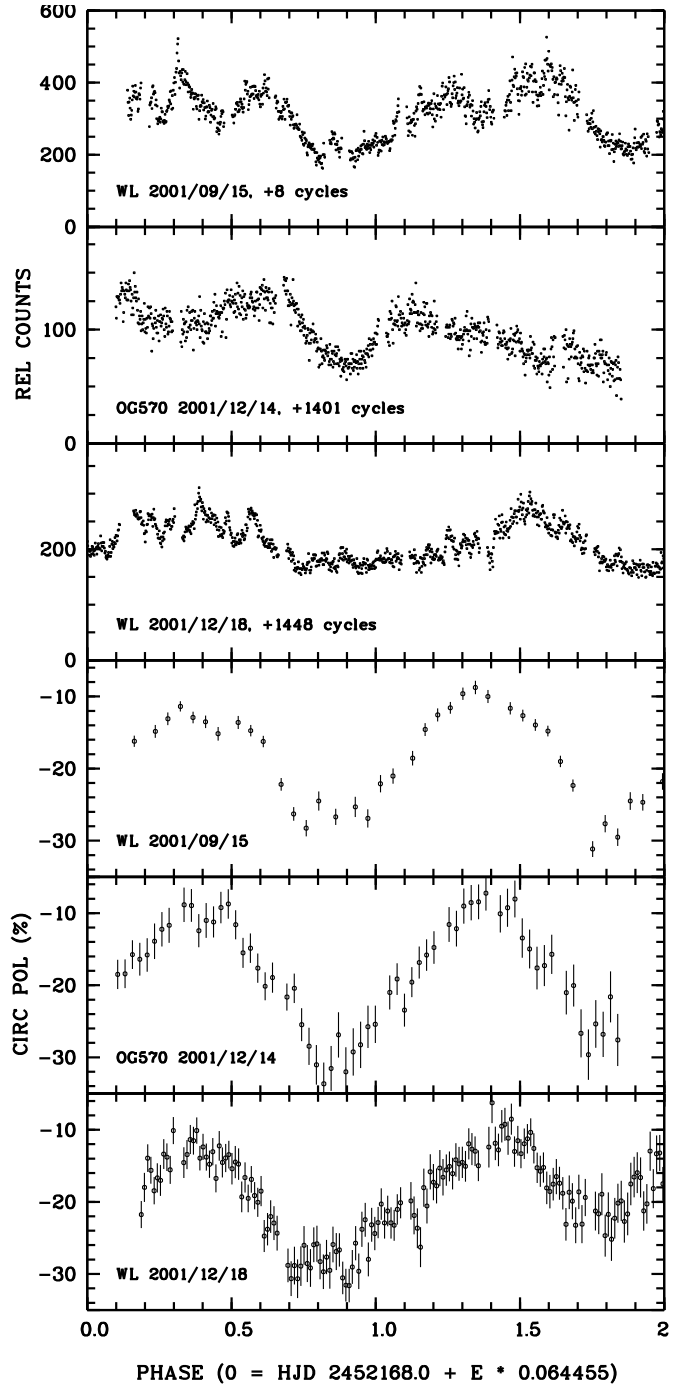}}
\caption{Photometry and polarimetry of RBS0206 with the SAAO 1.9m telescope.
Dates and filters used are indicated in the individual panels.
The top three panels show the photometry, the bottom 
panels the circular polarimetry. The data are phased according to
the ephemeris at the abscissa, but shown in original time sequence.\label{f:pol206}
}
\end{figure}

\section{New magnetic cataclysmic variables}
\subsection{RBS0206 (= 1RXS\,J012851.9$-$233931)}

The identification of RBS0206 as a magnetic cataclysmic variable of AM Herculis
type was announced by Schwope et al.~(1999). The identification 
was based on one low-resolution spectrum taken in a low accretion state 
and some occasional I-band photometry. The spectrum showed one pronounced
cyclotron line at $\sim$8000\,\AA\ probably originating in a low temperature 
accretion plasma, $kT \sim 2$\,keV, with a field strength of 45\,MG, and Zeeman shifted 
photospheric Balmer absorption lines indicating a mean magnetic field strength
of $\sim$36\,MG. The photometry suggested a periodicity of about 90 or 146 min, 
respectively.

\begin{figure*}[th]
\includegraphics[width=87mm,bbllx=37pt,bblly=100pt,bburx=462pt,bbury=766pt,clip=]{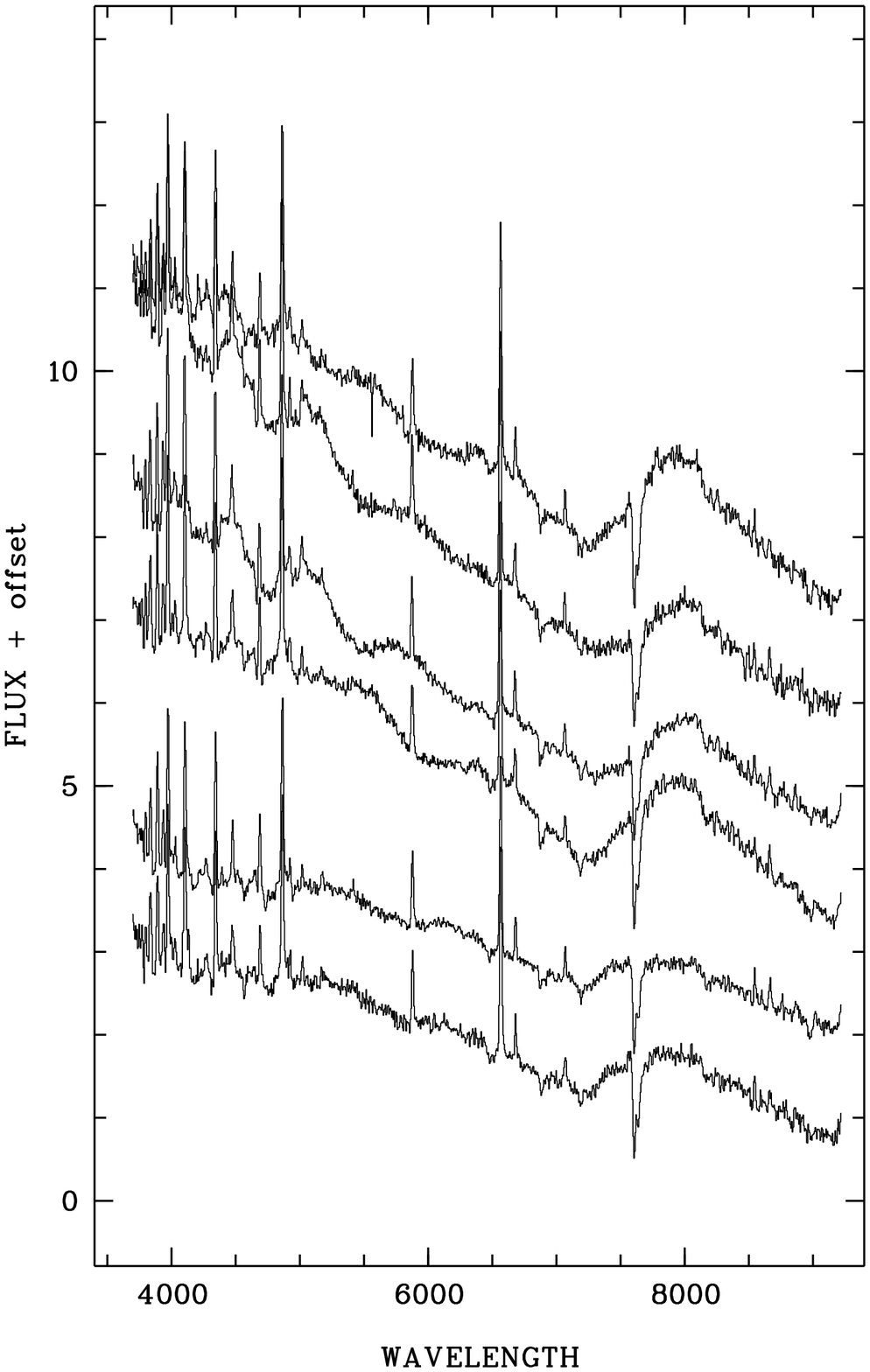}%}
\includegraphics[width=87mm,bbllx=37pt,bblly=100pt,bburx=462pt,bbury=766pt,clip=]{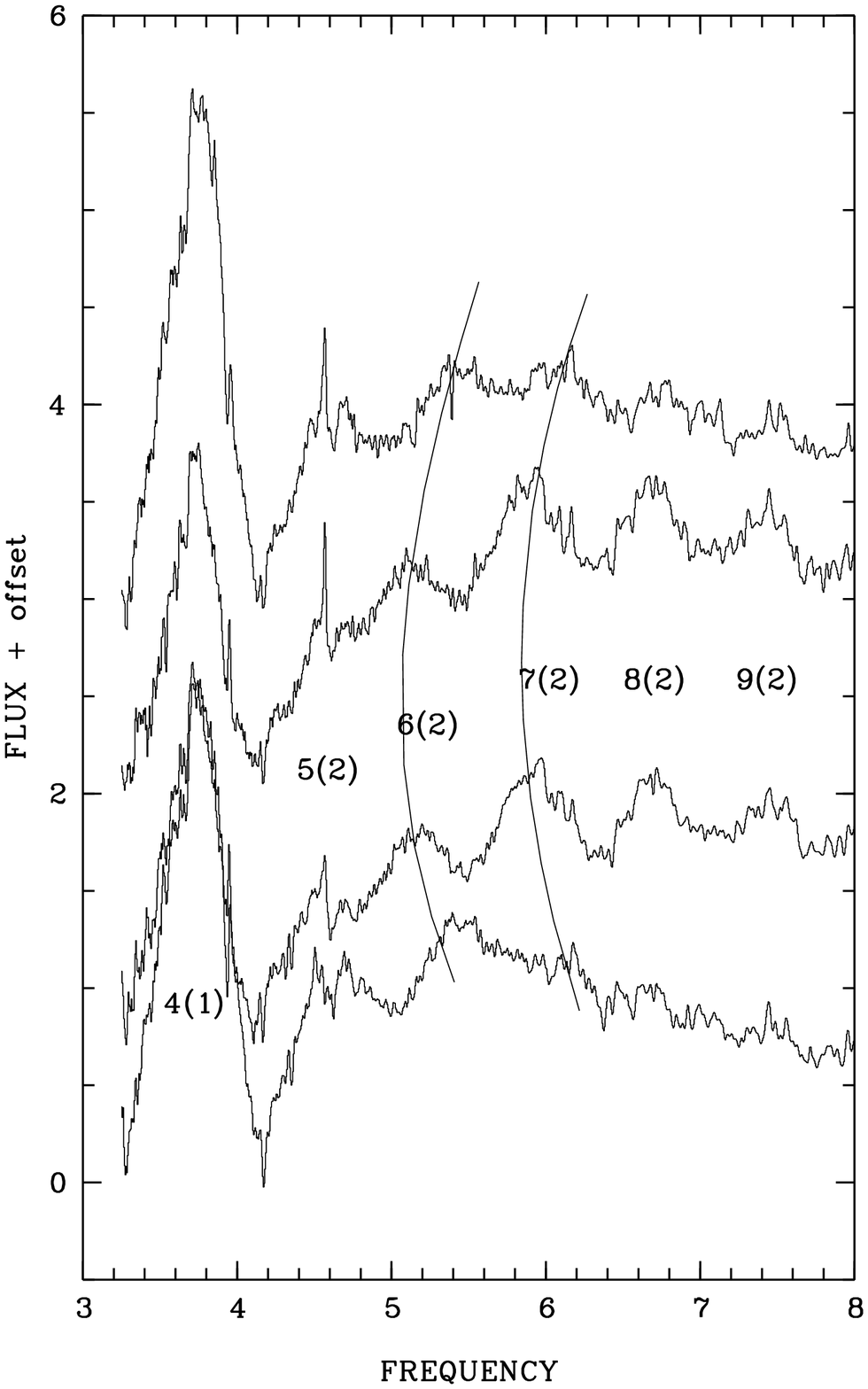}%}
\caption{{\it (left, a)} Low-resolution spectra around the spin cycle of RBS0206.
Flux units are $10^{-16}$\,erg\,cm$^{-2}$\,s$^{-1}$.
The individual spectra were plotted with an offset of 1.3 flux units
with respect to each other.
{\it (right, b)} The top four spectra of (a) after subtraction of the 
minimum spectrum, represented as $F_\nu$ vs.~$\nu$, plotted 
with an offset of 1 flux unit with respect to each other. 
Frequencies are given in $10^{14}$\,Hz.
Individual cyclotron humps of the two accretion spots are labeled by integer 
harmonic numbers.\label{f:r206mul}}
\end{figure*}

Follow-up photometry, polarimetry, and low-resolution spectroscopy was obtained
using the SAAO 1.9m telescope equipped with the UCT polarimeter in September
and December 2001, and with the ESO 3.6m telescope equipped with EFOSC2 in
December 2000. Results of the new observations are
shown in Figs.~\ref{f:s206} to \ref{f:r206cyc}.

In December 2000 a total of 9 spectra were taken with individual integration 
times of 900\,sec. The spectra cover the wavelength range 3700--9300\,\AA\ 
with a FWHM resolution of about 20\,\AA. The mean spectrum of the follow-up
observations is reproduced in Fig.~\ref{f:s206} in comparison with the 
discovery spectrum taken with the same instrumental setup at the same 
telescope. In December 2000 the source has returned to a high accretion 
state which is evident from the strong emission lines of Hydrogen, neutral 
and ionized Helium, the disappearance of the photospheric absorption lines
and the blue continuum. The low-state spectrum indicated a system 
brightness of $V \simeq 18\fm9$, in the high state of December 2000
the source varied between $V = 17\fm5$ and $18\fm0$.

Despite the large differences between the 
low and high state spectra, there is one striking similarity, the 
pronounced cyclotron line at 8000\,\AA.  This line is almost equally luminous
in both accretion states which is indicative of a low-temperature
accretion region persistently present irrespective of the 
mass accretion rate.

The photometric and polarimetric observations were used in order 
to derive the spin period of the white dwarf. Data were obtained 
in 5 nights between HJD 2452258 and 2452261 comprising about 1500
cycles of the binary. The light curves and circular polarization 
curves obtained in three of the nights are shown in Fig.~\ref{f:pol206}.
Filters and observation dates are indicated in the figure.
Linear polarization observations were also obtained but revealed no
clear detection of linearly polarized light.

Photometric variations by a factor of 2 are evident from the figure.
There seems to be a kind of on-off pattern present in the light curves 
with a pronounced photometric minimum at phase 0.9 (in filter OG570), 
the photometric minimum and the length of the bright phase, however,
are very unstable. The photometric variations did not yield 
reliable periods when period search algorithms were applied to these 
data.

The circular polarization signal displayed much smoother variation 
between $-10\%$ and $-30\%$ at any of the five occasions. The polarization
curves were found to be much more stable than the light curves 
itself and were used for a period search with the analysis-of-variance
method provided by the MIDAS data reduction package. A periodogram showed 
the most prominent peak at $f = 15.5147$ cycles per day, corresponding 
to a period of $92.815 \pm 0.003$\,min. This is accepted as the spin 
period of the white dwarf and likely to be the orbital period of the 
binary. Phases in Fig.~\ref{f:pol206} were calculated using this 
period with the zero point chosen arbitrarily at the beginning of the
photometric campaign. The accuracy of the polarimetric period is of
sufficient accuracy to connect the $\sim$1500 cycles covered
in 2001, but is insufficient to extend back to the earlier observations.

The high degree of the circular polarization is striking und underlines
the magnetic nature of RBS0206. The non-vanishing circular polarization 
indicates that the accretion spot where the cyclotron radiation 
originates never rotates behind the limb of the white dwarf but is
continously in view. The high degree of circular polarization 
at phase 0.8 -- 0.9 on the other hand suggests that we are almost 
looking pole-on at this phase. Both geometrical constraints together
indicate $i \simeq \delta \simeq 40\degr$ ($i$: orbital 
inclination, $\delta$: latitude of the field line in the accretion spot
with respect to the rotation axis).

The emission lines in the nine spectra obtained in December 2000
display radial velocity variations. When fitted with a sine curve,
the H$\alpha$ lines give a radial velocity amplitude of 125\,\kmps\ 
and a period of $96\pm4$\,min. The latter value is compatible with the 
polarimetric period and thus suggestive of $P_{\rm spin} = P_{\rm orb}$. 
Maximum recessional velocity occurs at
optical minimum. It is assumed, that the optical light curve 
is modulated mainly by the beaming properties of the cyclotron 
source. Optical minimum occurs, when the observer is looking pole-on.
The phasing of the radial velocity curve then implies that the 
bulk of the line radiation originates from the magnetically 
channeled accretion stream.

\subsubsection{Cyclotron spectroscopy of RBS0206}
The nine low-resolution spectra obtained in December 2000 were phase-folded
using the 92.8 min spin period. This resulted in spectra at 6 independent
phases. These are shown in 
Fig.~\ref{f:r206mul}a as $F_\lambda$ vs.~$\lambda$. 
The two lowest spectra were obtained in the 
photometric minimum. They still show the strong cyclotron 
emission line at 8000\,\AA\
found also in the low accretion state. Apart from this
line the minimum spectra do not show further pronounced  cyclotron humps.

The four spectra obtained in the bright phase, however, show a complete system 
of cyclotron humps at wavelengths shorter than 8000\,\AA. These lines show
a strong shift in wavelength as a function of phase, 
whereas the line at 8000\,\AA\ remains 
stationary. In the third spectrum from the top in Fig.~\ref{f:r206mul}a
the humps are centered at 6700, 5800, 5000, 4400, and 
4030\,\AA, respectively. This spectrum will be investigated in more
detail below.

%We are investigating the cyclotron spectrum of RBS0206 in two ways
%(see Figs.~\ref{f:r206mul}b and \ref{f:r206cyc}).
In order to underline the phase-dependent motion of the 
cyclotron lines we show the four bright phase spectra  
after subtraction of the average minimum spectrum in 
Fig.~\ref{f:r206mul}b ($F_\nu$ vs.~$\nu$) and connect two of the cyclotron 
harmonics with a curved line. The shift of the harmonics is of 
order of half the separation between two humps. It would
be quite natural to assume, that the additional cyclotron humps 
at short wavelengths are higher harmonics of the same source of 
radiation which emits the pronounced  line at 8000\,\AA. This
interpretation has severe difficulties, since the 8000\,\AA\
hump and those at shorter wavelengths cannot be explained with 
one common field strength. In addition, the 8000\,\AA\ hump 
remains stationary, whereas the other lines move with phase 
by several hundreds of \AA ngstroms (see Fig.~\ref{f:r206mul}b). 

\begin{figure}[th]
\resizebox{\hsize}{!}{
\includegraphics[bbllx=52pt,bblly=116pt,bburx=520pt,bbury=446pt,angle=-90,clip=]{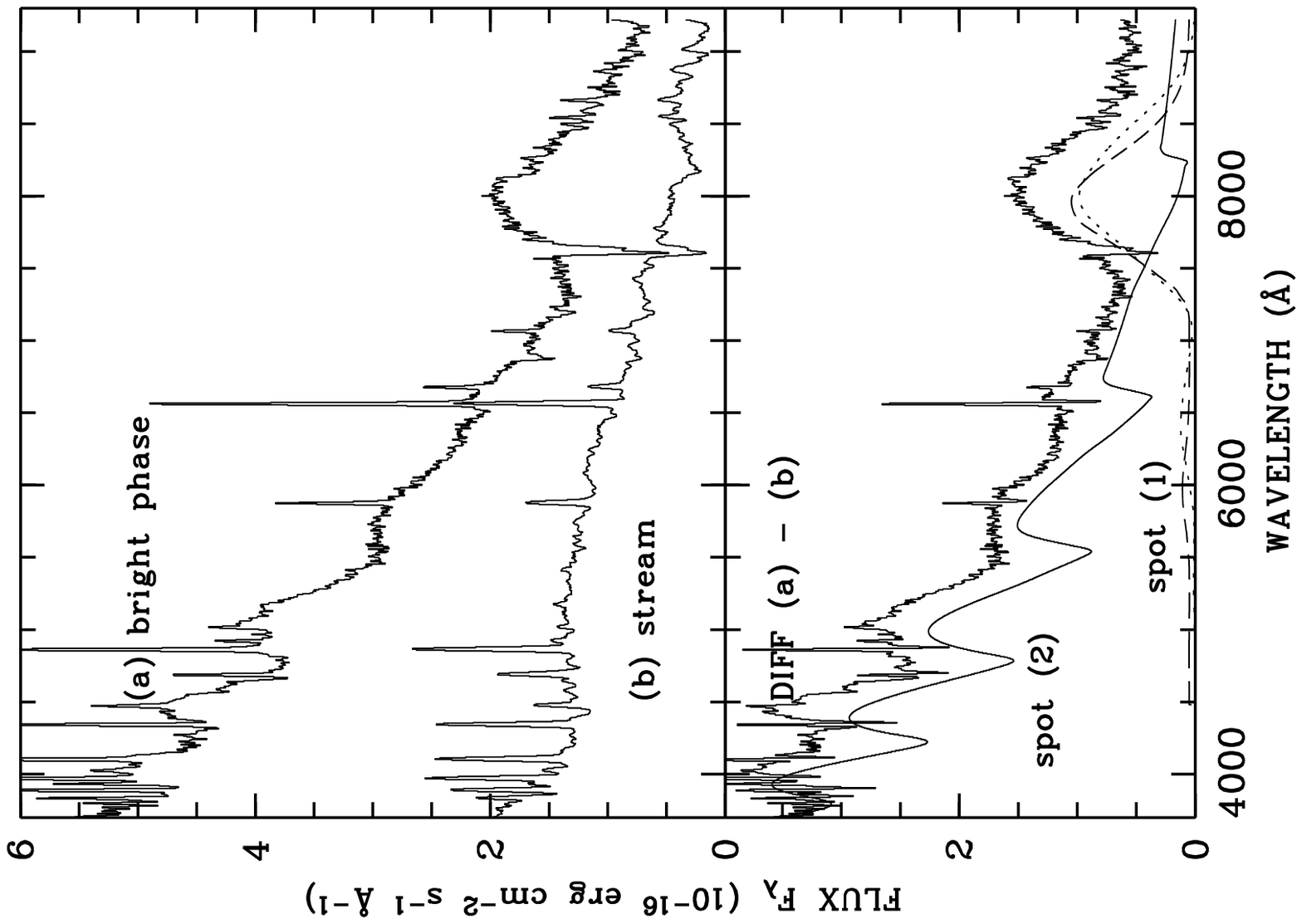}}
\caption{In the upper panel a high-state, bright-phase spectrum
of RBS0206 is shown. The spectrum below is the difference between the
orbital minimum in the high state and the single low-state spectrum 
obtained in the 1998 low state. In the lower panel the difference spectrum is 
shown, which is regarded as of pure cyclotron origin. The three cyclotron
models are explained in the text.
\label{f:r206cyc}
}
\end{figure}

We were thus forced to assume that the bright phase spectrum of
RBS0206 is fed by two independent sources of radiation.
Modeling of these two sources was performed for the 
bright-phase in the following manner.
We estimated the contribution of the accretion stream to the 
optical spectrum by assuming that the difference spectrum between 
optical minimum in the high state and the low-state discovery 
spectrum is solely due to emission from the stream. This 
spectrum is shown as the 
lower curve in the upper panel of Fig.~\ref{f:r206cyc}.
The upper curve is the average of the two 
high-state spectra which show the most pronounced
and most redshifted cyclotron lines ($2^{nd}$ and $3^{rd}$ from top
in Fig.~\ref{f:r206mul}). According to cyclotron 
theory this aspect corresponds to the maximum angle between the line 
of sight and the magnetic field in the emitting plasma. From 
polarimetry we estimated this angle to be of order 
$\Theta \simeq 80\degr$ (see above). 
The difference between the two spectra in the 
upper panel of Fig.~\ref{f:r206cyc} is regarded as pure 
cyclotron radiation and shown in the lower panel of that figure. 

We compare this spectrum with suitably scaled superpositions 
of homogeneous, isothermal cyclotron spectra. Since we are lacking 
sufficient phase resolution and phase coverage 
for a detailed comparison of the 
models with the observations, these shall be regarded as 
tentative only.

The model rising to the blue end of the spectrum was computed 
for $kT = 10$ keV, $\Theta = 80\degr$, $B = 32$\,MG. 
We refer to this model as spot(2) in Fig.~\ref{f:r206mul}
It is a superposition of three spectra with different plasma (density)
parameters $\Lambda$ and an optically thick Rayleigh-Jeans
part. The value of $\Lambda$ varies between $10^5$ and $10^8$.
This choice ensures that 
sufficient high harmonics remain optically thin in 
the individual spectra.
A rather high temperature of 10 keV was chosen for this part of the 
spectrum, since only at temperatures as high as this, do
harmonics move with aspect angle i.e.~spin phase 
(see Schwope 1990 for a general 
discussion of the motion of cyclotron humps). 
The general slope of the 
spectrum as well as the position of the cyclotron humps shortward
of 7300\,\AA\ are well described with this model. It fails to
reproduce the pronounced hump at 8000\,\AA, both in position 
and intensity. 

The dashed line in the lower panel of Fig.~\ref{f:r206cyc}
is the same model as shown in the discovery 
paper by Schwope et al.~(1999). It was computed for a temperature
of 2 keV, $\Theta = 50\degr$, $\Lambda = 10^2$, $B = 45$\,MG, and
could well reproduce the shape and position of the hump at 8000\,\AA.
Due to the very low value of $\Lambda$ it could also explain the 
non-detection of the next higher harmonic (the fourth), which was
completely optically thin and just too weak to be detectable.

The presence of two accretion spots at the same time, one with 32\,MG,
the other with 45\,MG, seems to be puzzling, as well as the extreme
low density of the spot which emits the 8000\,\AA\ hump.
We therefore explored whether the hump at 8000\,\AA\ could also be explained 
with plasma emission in a field of strength of about 32\,MG. 
A single low-temperature spectrum, $kT = 2$\,keV, $B = 33.7$\,MG,
indeed reveals a hump at the observed wavelength,
the  fourth harmonic of the cyclotron fundamental (indicated as 
this in Fig.~\ref{f:r206mul}b).
Such a line, however, is much too narrow in order 
to explain the cyclotron line as a whole. 

We therefore calculated a superposition 
of low-temperature, low-density, $kT = 2$\,keV, $\log\Lambda = 1$,
cyclotron models for a range of field strengths between 32 and 35.5 MG.
Such a model is shown as dotted line in the lower panel of Fig.~\ref{f:r206cyc}
and it reproduces the 8000\,\AA\ hump very well.
At the given low 
temperature, both models (33 and 45 MG) 
predict no considerable motion of the cyclotron humps as a function of phase.
We conclude that we cannot distinguish between the alternative 
models with the present data. 
The high-field model is attractive for its simplicity,
the low-field model is attractive for the field strength 
being close to the value 
used for modeling of the blue part of the spectrum. 

Both scenarios are possible, two spots at more or less 
the same location (field strength) on the white dwarf
or two spots with some separation, at least as far as the 
field strength is concerned. One region is rather dilute 
and persistently accreting during low and high accretion states, 
the other region whose density is in the normal range for 
polars, is accreting in the high state only. There is no 
firm clue on the accretion geometry apart from 
the likely combination $i \simeq \delta \simeq 40\degr$, 
which means that both accretion spots are continously in view.
In order to disentangle the accretion geometry one needs
to locate the secondary star in the orbit. Since the companion 
is not seen directly, the only way to achieve this goes
via the identification of an emission line of reprocessed
origin from the surface of the secondary in a high accretion 
state. Further observations, both spectroscopically and
polarimetrically, are highly demanded in order to unveil 
the secrets of this intriguing system.

\begin{figure*}[th]
\resizebox{\hsize}{!}
{\includegraphics[bbllx=95pt,bblly=57pt,bburx=480pt,bbury=760pt,angle=-90,clip=]{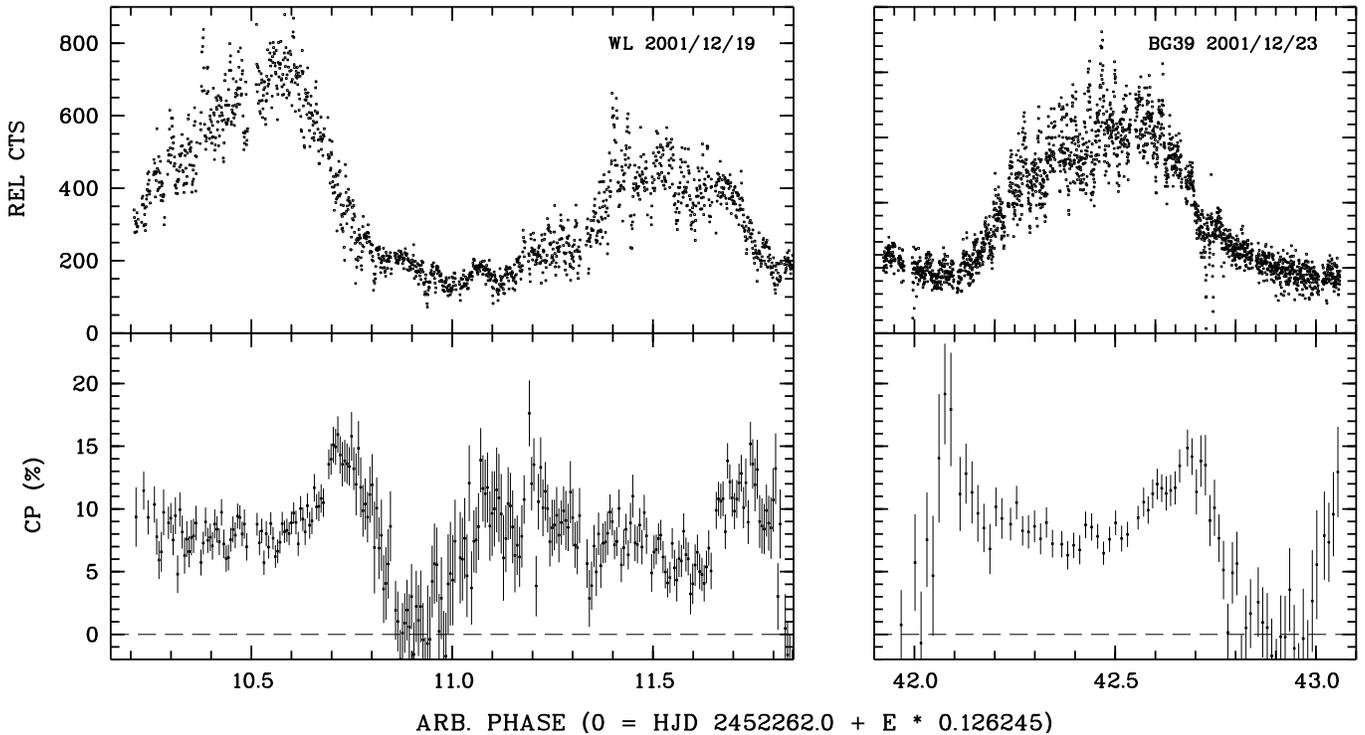}}
\caption{Polarimetry of RBS0324 obtained December 2001 with the 1.9m SAAO. Shown
are relative brightness (top) and the degree of circular polarization. Data are
shown in original sequence and were not phase-folded. The data on December 19 were 
taken in white light, those at December 23 through a blue BG39 filter.\label{f:pol324}
}
\end{figure*}

\begin{figure}
\resizebox{\hsize}{!}
{\includegraphics[bbllx=55pt,bblly=85pt,bburx=526pt,bbury=740pt,angle=-90,clip=]{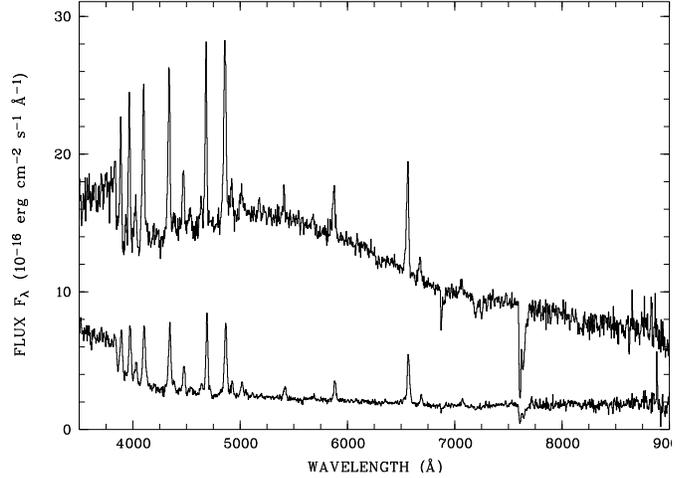}}
\caption{Identification spectra of RBS0324, 
obtained with the ESO 1.5m telescope on December 1997.
The two spectra represent the orbital minimum and maximum, respectively.\label{f:s324}
}
\end{figure}
\subsection{RBS0324 (= 1RXS\,J023052.9$-$684203)}
The identification spectra of \#324, obtained with the ESO 1.5m telescope, showed the typical 
line features of a high-accretion rate AM Herculis binary (polar), broad, asymmetric 
emission lines of H, HeI and HeII. Follow-up photometric, polarimetric, and 
spectroscopic observations 
were performed with the SAAO 1.9m telescope at several occasions (August 1998 --
December 2001), and the ESO/Dutch 90cm telescope (October 1998).
These observations consistently revealed a period of 181.8\,min, which we regard as the 
binary period of the system. The results of polarimetric observations obtained in December 2001
are shown in Fig.~\ref{f:pol324}. These data show the high degree of variability of the source
on timescales of minutes and hours. In its high state, the source varied between $V = 16^m$ and
18$^m$.
Polarimetry immediately proves the tentative identification as a polar through
the high degree of circular polarization, reaching $20\%$ at certain phases, 
modulated with the period of the binary.  

For future use, e.g.~the derivation of a long-term ephemeris, we
list in Table~\ref{t:r324max} the times of the centroids of the bright 
phase as obtained by Gaussian fits to the light curves 
at the different occasions in 1998 and 2001.
\begin{table}[h]
\caption{Times of the center of the bright phase of RBS0324}
\label{t:r324max}
\begin{tabular}{cl}
\hline
Date & $T_{\rm cent}$\\
YYMMDD & [HJD]\\
\hline
981029 & 2451115.6722\\
981031 & 2451117.5649\\
981101 & 2451118.8266\\
011219 & 2452263.3265\\
011219 & 2452263.4543\\
011223 & 2452267.3632\\
\hline
\end{tabular}
\end{table}

The light curve and polarimetric variation can be naturally explained 
by strong cyclotron beaming of one accreting pole, which is continously
in the observable hemisphere. At photometric maximum the angle between 
the observer and the magnetic field in the accretion spot is largest.
At photometric minimum we are observing almost from above the spot.
This should lead to a high degree of circular polarization at this phase.
Indeed an increase of the degree of polarization is observed at beginning
and end of the faint phase, in the center of the faint phase however the
signal becomes depolarized. This is likely due to depolarization
by background radiation (photospheric radiation from the stars and recombination
radiation from the accretion stream) and possibly due to absorption 
in the intervening accretion stream. The pattern of photometric 
and polarimetric variability resembles to a large extent that of MR Ser 
(Liebert et al.~1982) and we propose a similar accretion geometry. 
The orbital inclination is about $i \simeq 40\degr$, the co-latitude of the magnetic 
field $\delta$ in the accretion spot must be almost the same.
Such a geometry ensures a large variation of the projection angle
of the magnetic field in the spot, which in turn 
causes pronounced polarization variations. It also ensures, that 
the $i + \delta < 90\degr$, so that no circular polarization sign 
reversal occurs.

Low-resolution spectra at orbital minimum and orbital maximum are
shown in Fig.~\ref{f:s324}. The difference is regarded as cyclotron 
spectrum from the accretion spot(s). The difference spectrum is a
smooth function of the wavelength with maximum at $\sim$5500\,\AA.
Individual cyclotron harmonics cannot be resolved, so that no 
direct clue to the magnetic field strength can be derived. The
shape of the cyclotron spectrum is very similar to that of 
EF~Eri (Schwope 1991), which has a high temperature ($kT \simeq 20$\,keV), 
optically thick (plasma parameter $\Lambda \simeq 10^8$), low-field
accretion plasma ($B = 13$\,MG). We can readily assume similar plasma
parameters for RBS0324. Contrary to EF Eri, we cannot derive 
independent evidence for the low field in RBS0324 by Zeeman absorption 
lines, neither from the accretion halo nor from the white dwarf's photosphere
(\"{O}streicher et al.~1990).  

The spectra of Fig.~\ref{f:s324} display TiO absorption bands from the 
secondary star in the near infrared. The low signal to
noise ratio of the spectra prevents us from properly determining
the spectral type. At the orbital period of \#324, a Roche-lobe
filling dwarf star secondary has a spectral type dM4 
(see e.g.~the fit by G\"{a}nsicke et al.~(1995) to the secondary
in AM Her at the same orbital period). We determined the contribution
of the secondary to the spectral hump at 7520\,\AA\, to 20--25\%, which 
implies a distance to RBS0324 of about $d = 250$\,pc.

The small number of 70 photons was detected during the
RASS only. The RASS X-ray spectrum is very soft and compatible 
with one blackbody-like component only ($kT \simeq 30$\,eV).
Such a fit is similar to the soft component of many
AM Her stars (Beuermann \& Schwope 1994) including the
prototype (Paerels et al.~1994).

\begin{figure}
\resizebox{\hsize}{!}
{\includegraphics[bbllx=55pt,bblly=85pt,bburx=526pt,bbury=740pt,angle=-90,clip=]{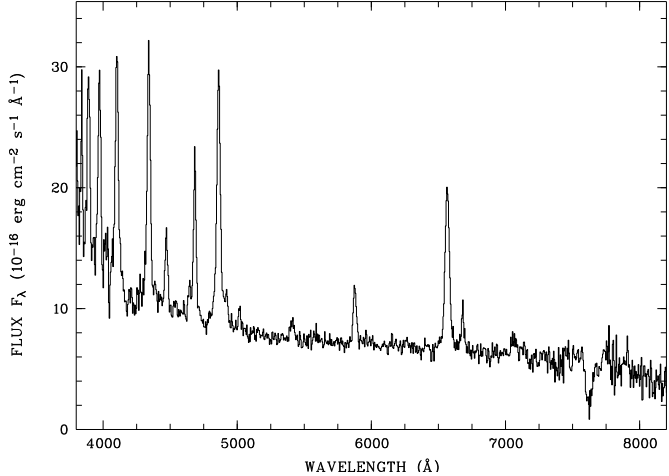}}
\caption{Discovery spectrum of RBS0541, obtained October 7, 1996, at Zelenchukskaya.\label{f:s541}
}
\end{figure}
\begin{figure}
\resizebox{\hsize}{!}
%{\includegraphics[bbllx=52pt,bblly=74pt,bburx=546pt,bbury=781pt,angle=-90,clip=]{lcs_541.ps}}
{\includegraphics[clip=]{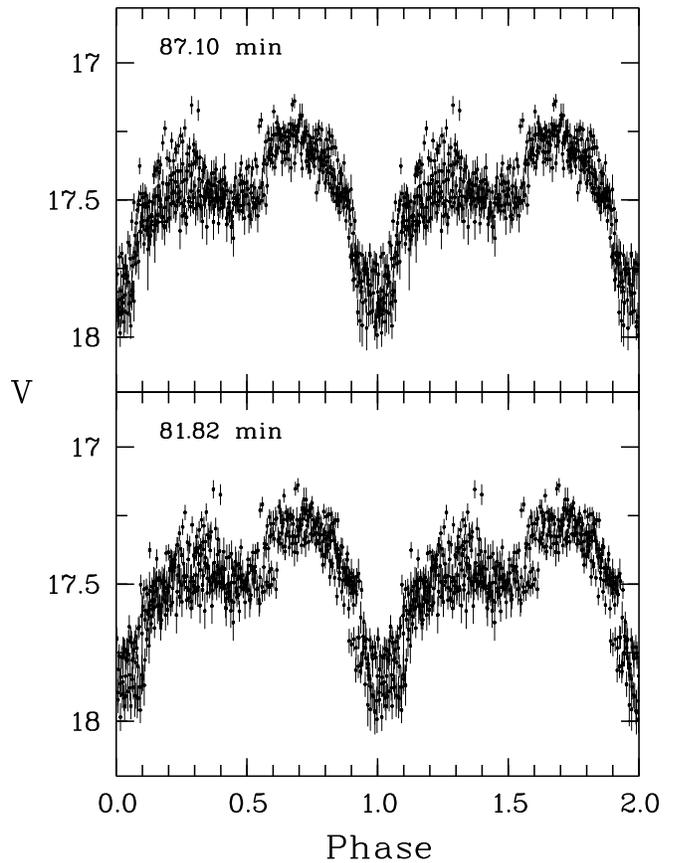}}
\caption{CCD-photometry of RBS0541, folded over the possible alias periods (observation dates
October 28 -- 31, 1998). Shown are differential V-band magnitudes 
as a function of orbital phase. All data are shown twice for clarity.\label{f:phot541}
}
\end{figure}

\subsection{RBS0541 (= 1RXS\,J042555.8$-$194534)}
The low-resolution discovery spectrum of RBS0541, 
taken at Zelenchukskaya, shows a smooth blue continuum and
asymmetric emission lines of H, HeI, and HeII. There is no obvious sign of the secondary star 
in the spectrum. The below-slit magnitude of \#541 at October 7, 1996, was $V = 16.7$, similar 
to the brightness listed in the USNO-A2 catalogue. Attempts were made to obtain further 
spectra with the 1.9m telescope at SAAO on August 20 and 23, 1998, and January 22, 1999.
These attempts failed due to the faintness of the system at these occasions, indicating
a deep low state of the binary. 

Time-resolved CCD-photometry with the ESO/Dutch 90cm telescope at La Silla was obtained
in the nights Oct 28 -- 31, 1998, through a $V$-filter.
Source intensities were determined using the DoPhot package 
(Mateo \& Schechter 1989) and intensities were reduced to the brightness of the 
$14\fm77$ comparison star USNO A2 U0675\_01566006 at 
RA = 4:26:00.41, DEC = --19:45:57.4. The source showed an overall 
brightness decline by $\sim$0.4 mag during those days. Apart from 
this, RBS0541 displayed pronounced periodic 
intensity variations with an amplitude of $\sim$0.8\,mag between 
$V \simeq 17.2$ and $\sim$18.0 (Fig.~\ref{f:phot541}). 
A period search with the analysis of variance technique 
implemented in MIDAS yielded a likely period of 87.1\,min with 
a possible alias period of 81.8\,min. The folded, de-trended light 
curves for both possible periods are shown in Fig.~\ref{f:phot541}.
Times of optical minima in RBS0541 are 2451115.8014,
2451115.8608, 2451116.7691, 2451117.6761, and 2451118.6447.

The RASS X-ray spectrum contains 77 photons.
It shows the typical shape of an X-ray spectrum of a polar with a hard
bremsstrahlung-like tail and a soft blackbody component.

The system shows all the hallmarks of a polar system, an emission 
line spectrum with asymmetric emission lines, also from high ionization 
species, a two-component X-ray spectrum and changes between high 
and low states. The periodic modulation of the optical 
light curve is interpreted as orbital period. At a period of 82--87\,min
the spectral type of the secondary must be later than M6. 

The shape of the light curve is similar to that of e.g.~V834 Cen,
indicating a similar accretion geometry 
with the main accretion spot being always in the visible 
hemisphere. The orbital minimum in such a geometry 
is caused by cyclotron beaming and stream occultation
and occurs, when the observer looks almost perpendicular
onto the accretion spot. The secondary minimum occurs due
to a partial self-eclipse of the accretion spot behind the 
limb of the white dwarf.
The likely orbital inclination is in the range 30\degr -- 60\degr,
the co-latitude of the accretion spot conversely in the range 
60\degr -- 30\degr.

\begin{figure}
\resizebox{\hsize}{!}
{\includegraphics[bbllx=55pt,bblly=85pt,bburx=526pt,bbury=740pt,angle=-90,clip=]{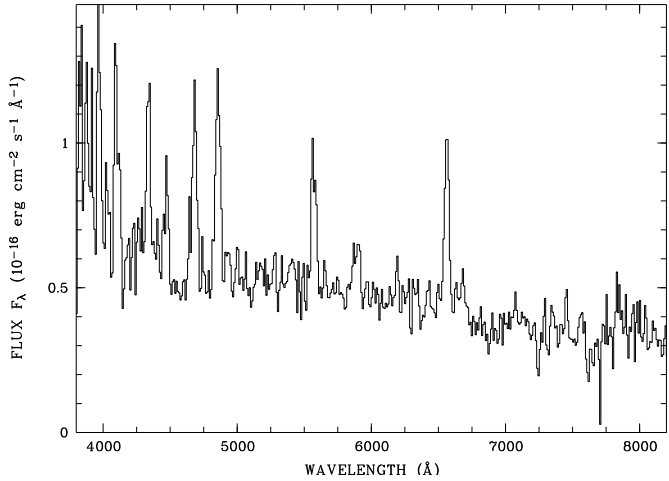}}
\caption{Identification spectrum of RBS0696 taken with MOSCA at the Calar Alto 
3.5m telescope on March 16, 1999.\label{f:s696}
}
\end{figure}

\begin{figure}
\resizebox{\hsize}{!}
{\includegraphics[bbllx=181pt,bblly=72pt,bburx=499pt,bbury=628pt,angle=-90,clip=]{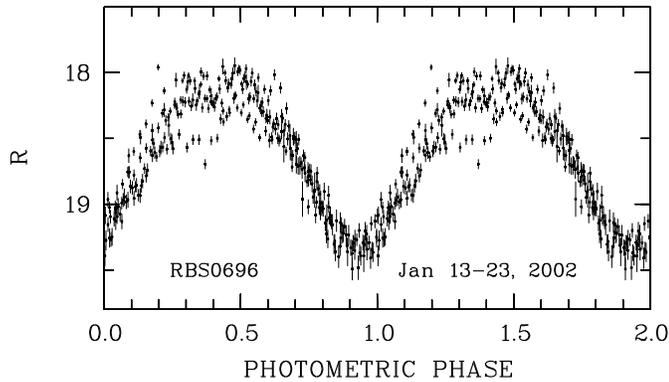}}
\caption{R-band photometry of RBS0696 performed with the Calar Alto 
1.23m telescope. The data were phase-folded (not averaged) over the 
period of 99.38\,m. \label{f:p696}
}
\end{figure}

\subsection{RBS0696 (= 1RXS\,J082051.2$+$493433)}
Source \#696 was identified as cataclysmic variable also by Cao et al.~(1999) within 
their identification programme of sources with high ratio between RASS X-ray and optical (DSS)
flux. Our identification spectrum of the $V=19\fm5$ optical counterpart, a 15min exposure 
taken with the 3.5m telescope at Calar Alto, Spain,
equipped with the MOSCA instrument on March 16, 1999, is reproduced 
in Fig.~\ref{f:s696}. It shows H, HeI, and HeII emission lines, respectively, which
are superposed on a blue continuum. In the spectrum shown by Cao et al.~mainly 
Balmer emission lines can be recognized. The emission lines in our spectrum 
have a larger equivalent width and the spectrum contains species with higher 
ionization potential. This is indicative of an accretion rate change.
No obvious spectral signature from the 
secondary star could be found. With the spectral resolution given it is 
impossible to judge, whether the emission lines are asymmetric or have 
multiple components. 

Photometric observations using the 1.23m telescope at Calar Alto were performed
during five nights between January 13 at 23, 2002. Differential $R-$band
magnitudes were derived with respect to the comparison star U1350\_07362511 
(USNO A2), which has $R=15\fm3$. Smooth, quasi-sinusoidal variations between 
$R = 18.0 - 19.4$ became evident 
with a period $(99.38 \pm 0.03)$\,min. No other period was found and we regard
$P = 99.38$\,min as the orbital period of the binary. The phase-folded 
data are shown in Fig.~\ref{f:p696}. It shows a large scatter by
$\sim$0.5\,mag at optical maximum and negligible scatter at optical 
minimum. This property is typical of polars and caused by irregular
accretion at the hot spot. The bright phase was centered on HJD
2452287.7318, 2452289.4597, 2452289.5283, 2452289.5979, 2452290.4224,
2452297.6691, 2452297.7409, respectively.

The softness of the X-ray spectrum, the high ionization 
emission line spectrum, an the presence of only one period, the likely 
binary period, make the identification as magnetic CV of 
AM Herculis type very likely. 

\begin{figure}
\resizebox{\hsize}{!}
{\includegraphics[bbllx=55pt,bblly=85pt,bburx=526pt,bbury=740pt,angle=-90,clip=]{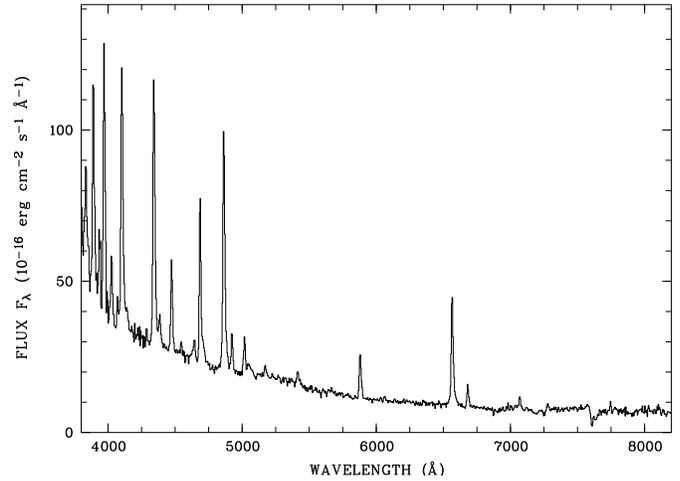}}
\caption{Discovery spectrum of RBS1563, obtained May 19, 1996, at Zelenchukskaya.\label{f:s1563}
}
\end{figure}

\subsection{RBS1563 (= 1RXS\,J161008.0$+$035222)}
RBS1563 was initially observed with the SAO 6m telescope at Zelenchukskaya, were the 
discovery spectrum was taken (Fig.~\ref{f:s1563}). The spectrum shows the typical 
hallmarks of a magnetic CV, a blue continuum, broad asymmetric emision lines 
of H, HeI and HeII, and some indication of the secondary star in the near infrared 
spectral range. These properties are similar to those reported in the 
single spectrum shown by Jiang et al.~(2000).
 
During the RASS a total of 207 photons were collected. X-ray variability by $\sim$100\%
is clearly present, reaching a peak count rate of about 1.4 cts/s. The resulting 
spectrum clearly contains two components, a soft blackbody-like component with 
$kT \simeq 30$\,eV, and a hard bremsstrahlung-like component (fixed at 20\,keV for 
a spectral fit). Although prominent, the soft component was not as dominant as in 
other magnetic CVs, which lead to a moderate ROSAT hardness ratio.

Follow-up spectroscopy and photometry took place using the ESO/Danish 1.5m telescope
and the AIP 70 cm telescope. The rather large observational body thus collected
deserves a separate publication (Schwarz et al., in preparation), 
and we give a very brief summary here.

RBS1563 is a magnetic CV of AM Herculis type, displaying high and low accretion 
states. The orbital period is 187.7 min. In the high state, 
the optical light curve is roughly 
sinusoidal, which is suggestive of a one-pole accretion geometry and a
moderate orbital inclination, $i = 30\degr - 60\degr$. 

\begin{figure}
\resizebox{\hsize}{!}
{\includegraphics[bbllx=55pt,bblly=85pt,bburx=526pt,bbury=740pt,angle=-90,clip=]{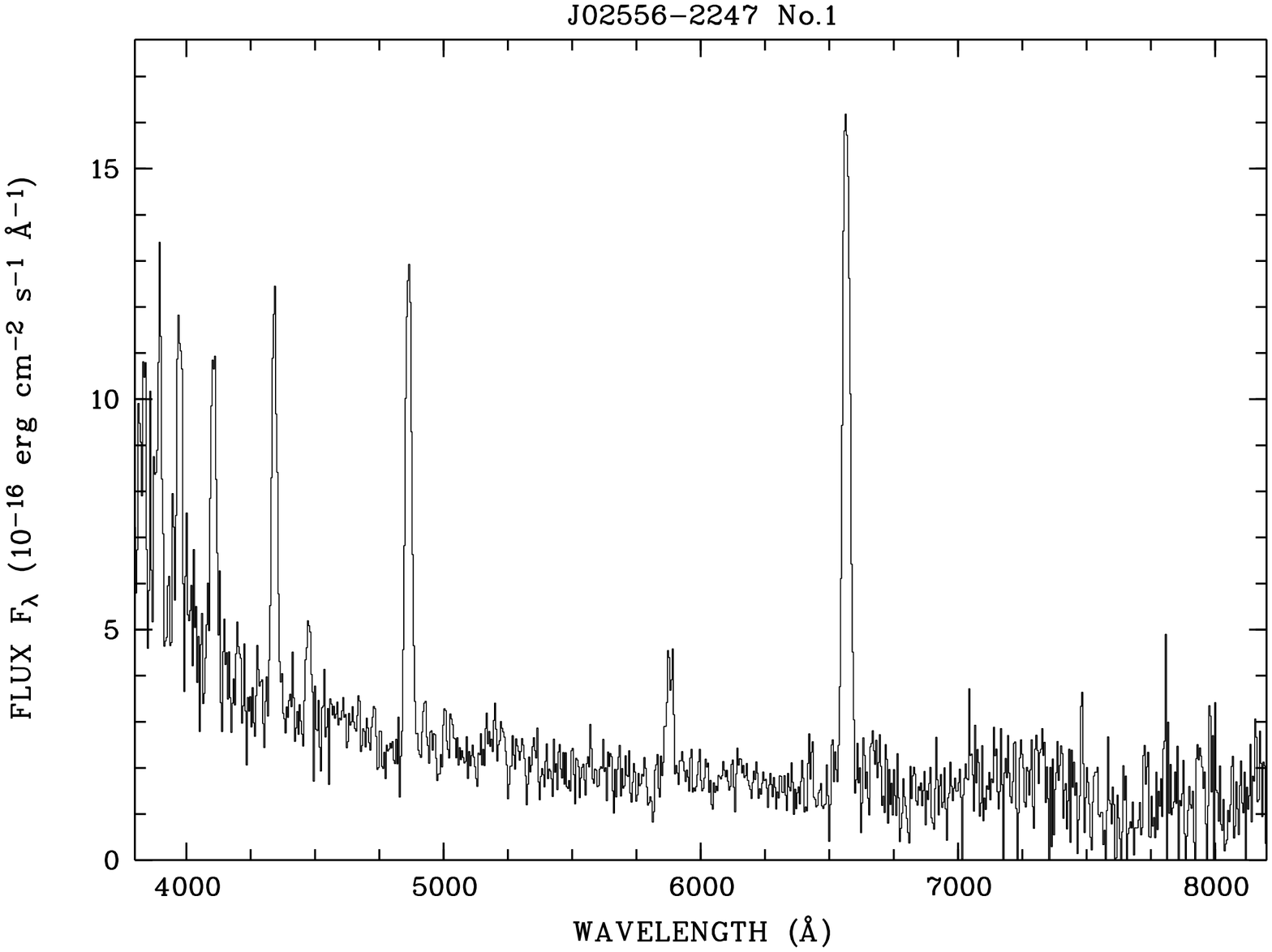}}
\resizebox{\hsize}{!}
{\includegraphics[bbllx=55pt,bblly=85pt,bburx=526pt,bbury=740pt,angle=-90,clip=]{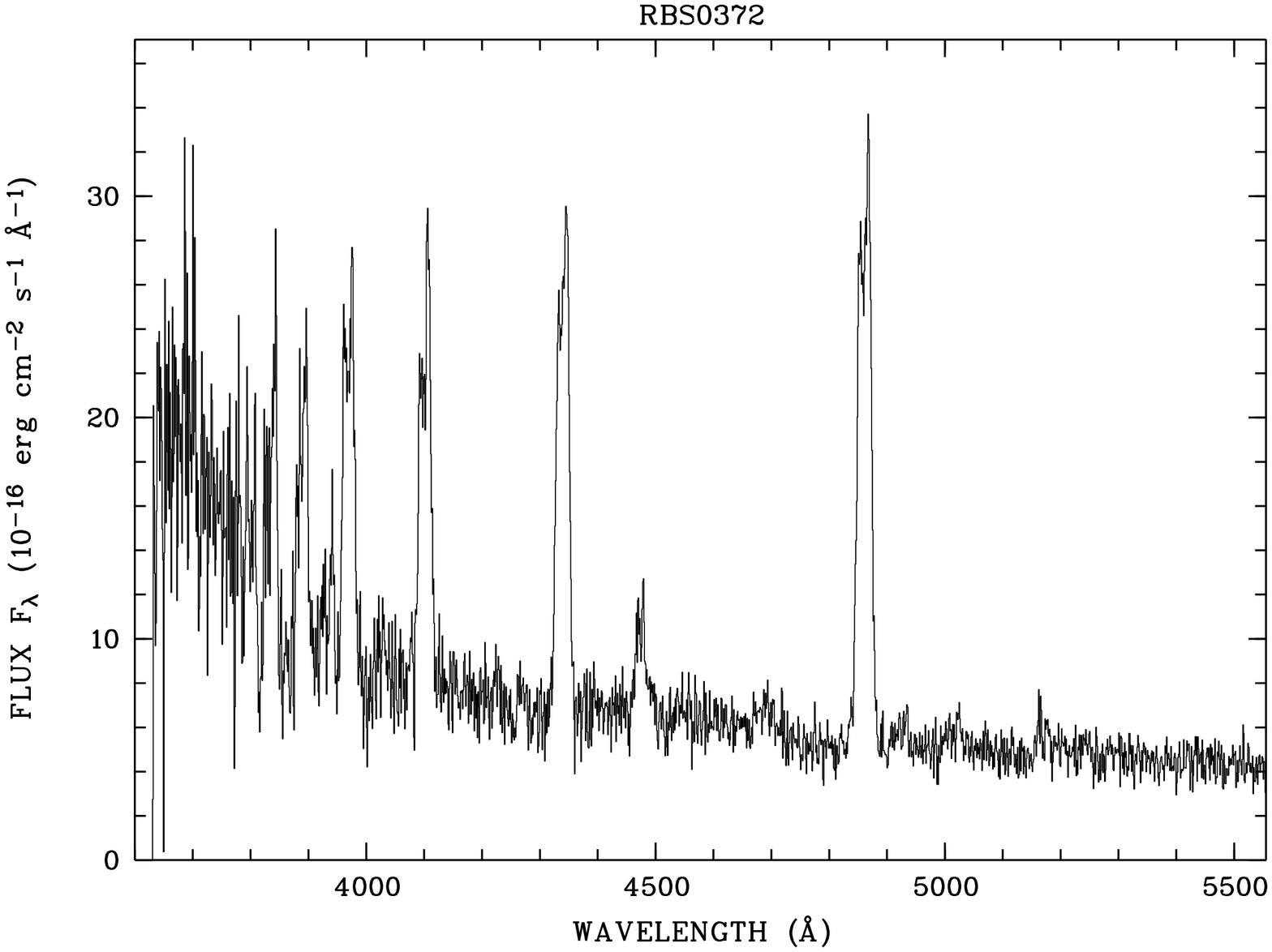}}
\caption{{\it (top)} Discovery spectrum of RBS0372 obtained October 7, 1996, at Zelenchukskaya.
{\it (bottom)} Spectrum with moderate resolution of RBS0372 obtained August 23, 1998, at SAAO.\label{f:s372}
}
\end{figure}

\begin{figure}
\resizebox{\hsize}{!}
{\includegraphics[bbllx=55pt,bblly=85pt,bburx=526pt,bbury=740pt,angle=-90,clip=]{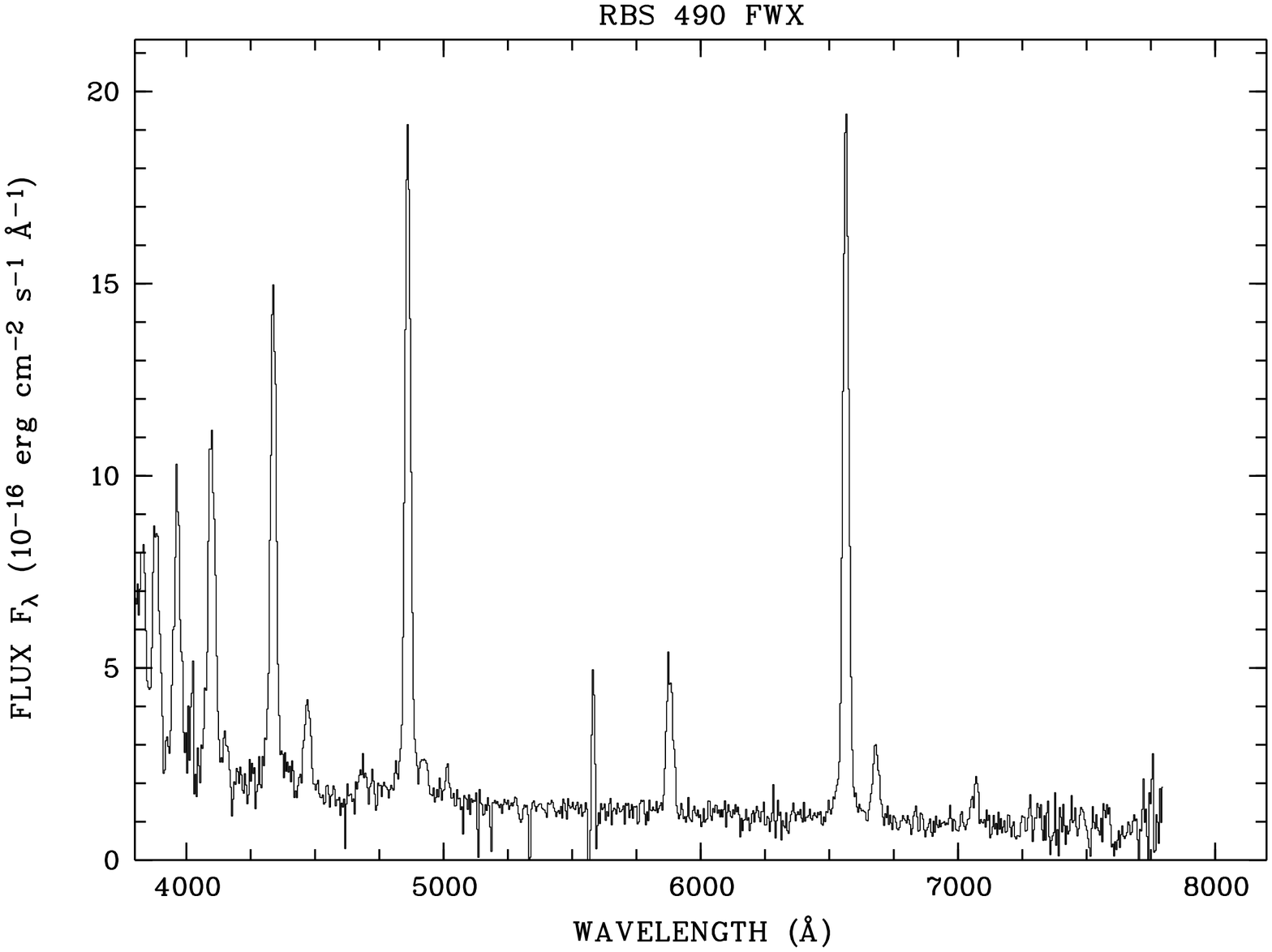}}
%}
%\end{figure}
%\begin{figure}[h]
\resizebox{\hsize}{!}
{\includegraphics[bbllx=55pt,bblly=85pt,bburx=526pt,bbury=740pt,angle=-90,clip=]{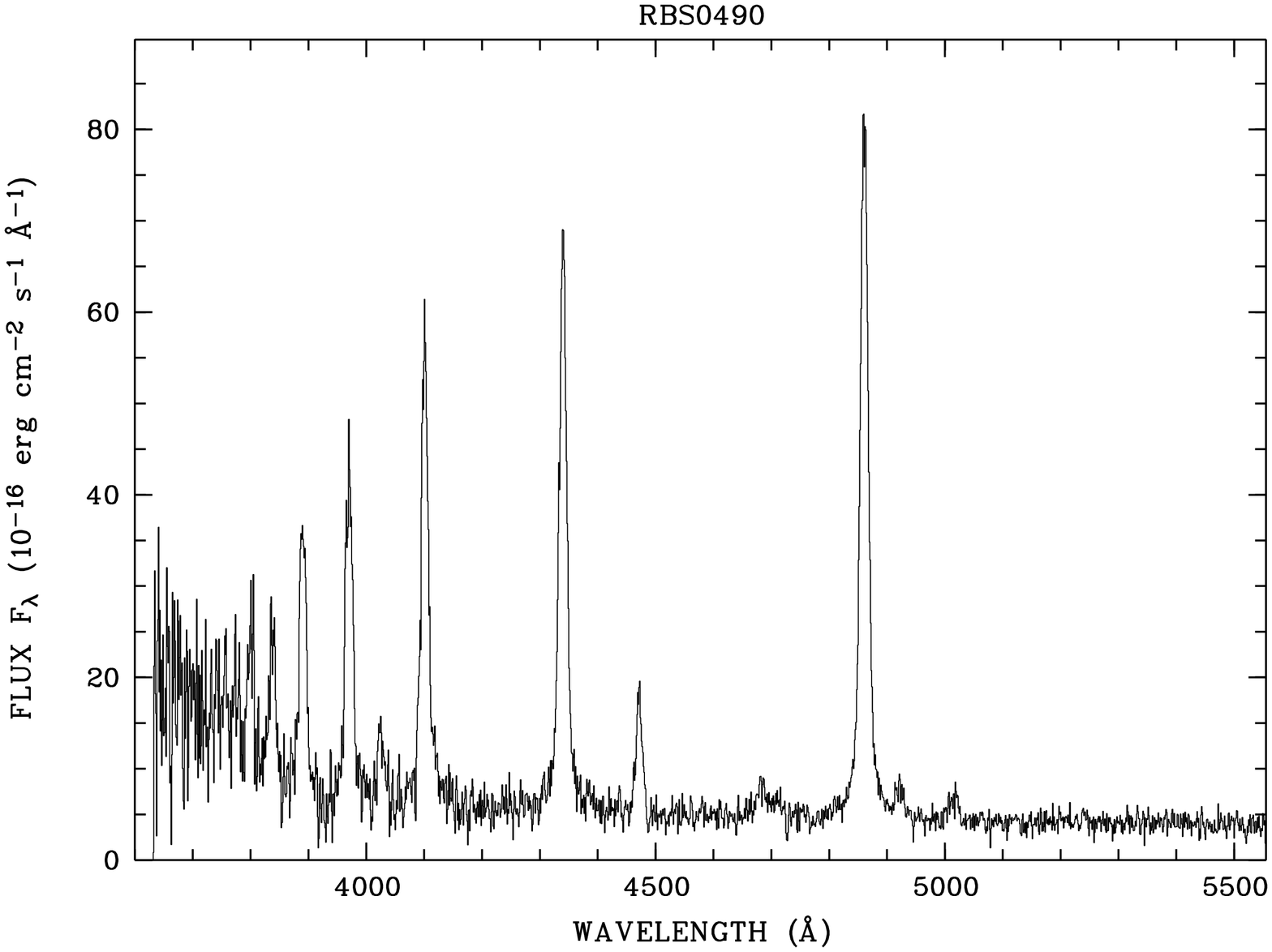}}
\caption{{\it (top)} Discovery spectrum of RBS0490, obtained October 10, 1997, at Zelenchukskaya.
%\caption{
{\it (bottom)} 
Mean blue spectrum of RBS0490, obtained on August 23, 1998, at SAAO.\label{f:s490}
%\label{f:s490saao}
}
\end{figure}

\section{New non-magnetic cataclysmic variables}

\subsection{RBS0372 (= 1RXS\,J025538.2$-$224655)}
The discovery spectrum of RBS0372, obtained on October 7, 1996, with the SAO 6m-telescope
is shown in Fig.~\ref{f:s372}. The derived below-slit magnitude was $V = 18\fm1$. It shows
pronounced  H-Balmer and neutral HeI emission lines. Emission of HeII was weakly detected.
There is no clear indication of the secondary star. Some spectra with higher resolution 
in the blue and the red spectral range were obtained with the SAAO 1.9m telescope.  
The mean blue spectrum taken on August 23, 1998, 
with effective exposure of 38\,min is reproduced also in Fig.~\ref{f:s372} (bottom panel).
It shows the same emission lines as in the discovery spectrum but this time
clearly resolved into a double-peaked profile. 
This is highly reminiscent of emission originating from an accretion disk 
with rather high inclination. The implied brightness was somewhat higher, 
$V \sim 17\fm6$, during the SAAO observations.
Time-resolved CCD-photometry performed in January 1997 in white light with 
10 sec time resolution shows pronounced modulation of 
the optical light curve by up to 0.5 mag. Those brightness variations happened 
on time scales of minutes, hours and nights, but without clear periodicity.
The relatively hard X-ray spectrum, the double-peaked emission lines and
the rather stochastic intensity variations led to our classification
as a non-magnetic cataclysmic variable, most probably a 
dwarf-nova in quiescence.

\subsection{RBS0490 (= 1RXS\,J035410.4$-$165244)}
The ROSAT X-ray hardness ratio HR1 = +0.57 of RBS0490 is the highest among all CVs
discussed here. A spectrum constructed from the 87 photons detected in the RASS 
can be fit with a 20 keV bremsstrahlung component. 
The identification spectrum of the $V\sim16^m$ counterpart (Fig.~\ref{f:s490}, top), 
obtained October 10, 1997, 
at Zelenchukskaya, shows pronounced H-Balmer and HeI emission lines and weak HeII 
emission lines. It has a flat, blue continuum. 
The low-resolution spectrum of \#490 is very similar to that of \#372,  
although with a somewhat smaller Balmer decrement. Further spectra with
moderate resolution were taken on August 23, 1998, at the Sutherland site of SAAO.
The mean combined spectrum (four exposures) which has effective integration 
time of 46\,min is reproduced also in Fig.~\ref{f:s490} (bottom).
The lines are resolved and double-peaked.
The optical and X-ray spectra are typical for a dwarf-nova in quiescence with 
a moderate to high inclination.

\subsection{RBS1411 = RHS40 (= 1RXS\,J143703.5$+$234236)}

\begin{figure}
\resizebox{\hsize}{!}
{\includegraphics[bbllx=55pt,bblly=85pt,bburx=526pt,bbury=740pt,angle=-90,clip=]{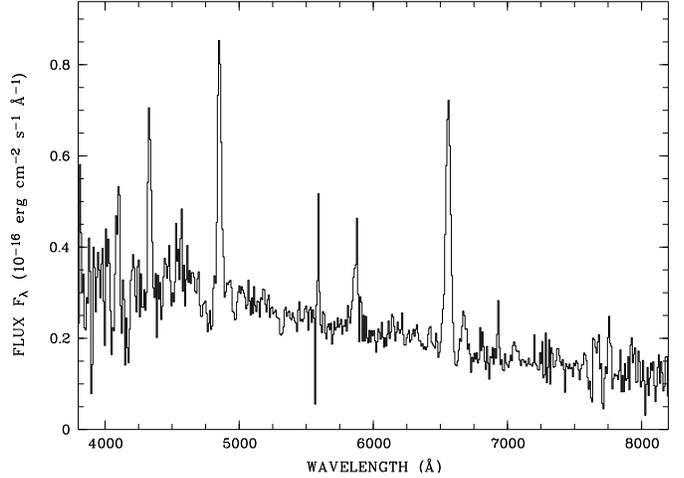}}
\caption{Discovery spectrum of RBS1411, obtained April 29, 1995, with the ESO/MPG 2.2m telescope
at La Silla, equipped with EFOSC2.\label{f:s1411}
}
\end{figure}

RBS1411 was identified already in the 
initial program aiming at the identification of a bright, hard subsample of
the RASS, the ROSAT Hard Survey (RHS). 
It was listed as number \#40 in the initial RHS-catalog (Fischer et al.~1998).
It has a hard RASS X-ray spectrum with HR1 = 0.57, compatible with a single-component
bremsstrahlung spectrum of $kT \simeq 10$\,keV. Further information on \#1411
is sparse. On the DSS the counterpart has an optical brightness of $V=19\fm1$.
It was similarly faint on April 29, 1995, when the identification spectrum 
was taken (Fig.~\ref{f:s1411}). Our spectrum is similar to that shown 
by Jiang et al.~(2000) which confirms our analysis.
The spectrum has a blue, smooth continuum with emission lines of Hydrogen 
and neutral Helium superposed. The flux decrease towards the blue end of 
the spectrum is probably not real, but due to an imperfect instrumental 
response function. Due to the absence of high ionization species and
the hard X-ray spectrum we classify this object as a non-magnetic CV, possibly 
a dwarf nova in quiescence (at the time of the optical observations). 
This is supported
by the X-ray to optical flux ratio $F_X/F_{\rm opt} = 30$ which is found 
to be unusually high 
($F_X (0.1 - 2.4 {\rm keV})= 2.5 \times 10^{-12}$\,erg\,cm$^{-2}$\,s$^{-1}$, 
$F_{\rm opt} = 10^{-0.4 V - 5.44} = 
8.3 \times 10^{-14}$\,erg\,cm$^{-2}$\,s$^{-1}$). 
This is the highest value among all non-magnetic CVs in the RBS, at the 
extreme end of the DN population and two orders of magnitude higher 
than the nova-like variables
(for comparison see the synoptic paper by Beuermann \&Thomas 1993, their Fig.~2).

\subsection{RBS1955 (= 1RXS\,J230949.6$+$213523)}

\begin{figure}
\resizebox{\hsize}{!}
{\includegraphics[bbllx=55pt,bblly=85pt,bburx=526pt,bbury=740pt,angle=-90,clip=]{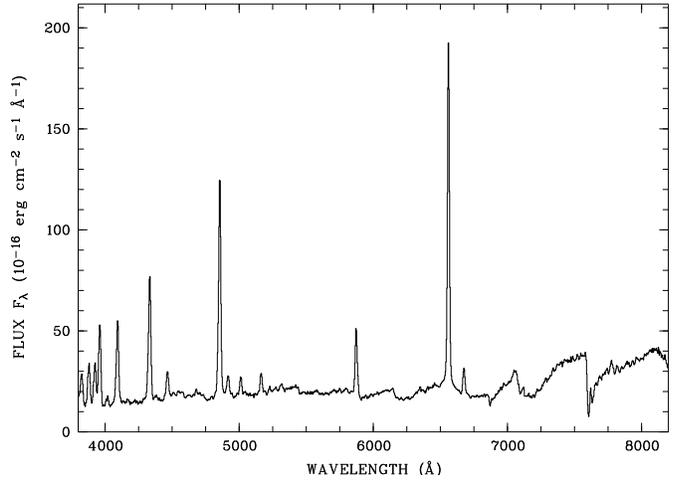}}
\caption{Discovery spectrum of RBS1955, obtained December 2, 1997, with the 1.5m telescope
at ESO, La Silla.\label{f:s1955}
}
\end{figure}

\begin{figure}
\resizebox{\hsize}{!}
{\includegraphics[bbllx=225pt,bblly=102pt,bburx=474pt,bbury=574pt,angle=-90,clip=]{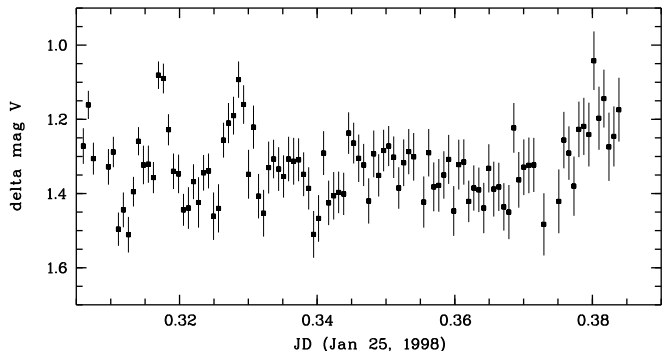}}
\caption{CCD-photometry of RBS1955 obtained January 25, 1998, at the Babelsberg site. 
Plotted are V-band differential magnitudes with respect to the comparison star 
26 arcsec WNW to the source.   \label{f:phot1955}
}
\end{figure}

With $V \simeq 15\fm6$, source \#1955 is by far the brightest of the new 
interacting binaries at optical wavelengths. The identification spectrum 
taken with the ESO 1.5m telescope at La Silla is reproduced in Fig.~\ref{f:s1955}. 
A spectrum similar to ours was reported also by Wei et al.~(1999).
They classified the system as CV without further specification.
The continuum is dominated by emission from an M-type star. 
Strong emission lines of H-Balmer and HeI are superimposed. 

In the RASS 101 X-ray photons were collected. 
X-ray variability is evident on a 2$\sigma$
level. Explanation of the RASS X-ray spectrum requires two radiation 
components, a soft blackbody-like component with $kT_{\rm bb}=38$\,eV, 
and a hard bremsstrahlung-like component which we fixed at 20\,keV.

Optical variability with an amplitude of $\sim$0.3\,mag was detected in 
differential V-band photometry, performed with the 70cm telescope of the AIP
at two occasions in 1998 (see Fig.~\ref{f:phot1955} for an example). 
%Photometric variations by a factor $\sim$2 on a timescale of 20\,min 
%were also observed in a sequence of three spectra taken in December 1997, 
%but only at the very blue end of the spectral range covered (below 4000\,\AA). 
A sequence of three spectra taken in the night of discovery with a time
resolution of 10 min does not show any variablity in the emission line 
pattern, neither in flux nor in position. 

A spectral classification using the schemes by Kirkpatrick et al.~(1999)
and Mart\'{\i}n et al.~(1999) revealed a spectral type of M3. 

On the basis of the optical spectrum with its strong M-type stellar
component, we tentatively classified this object as symbiotic binary
(Schwope et al.~2000).
The possible short-timescale variability does not contradict this
classification (see Sokoloski et al.~2001 for a comprehensive 
optical variability study of symbiotics). 
If this classification would be correct, RBS1955 would
only the third symbiotic in the RBS, the others being AG Dra and 
St\,Ha32, which both display supersoft spectra. \#1955 also has a 
soft component, but there is a hard component 
present in addition. 

The classification as a symbiotic binary, however, is far from being 
unique. As U.~Munari kindly pointed out to us, the complete absence
of high ionization lines, the rather high HeI triplet to singlet
ratio 5876/6678 and the rather flat Balmer decrement are not 
compatible with this tentative classification. 
The absence of a strong NaI~D absorption at 5893\,\AA\ 
seems to exclude a dwarf classification
for the M star (which would suggest a normal CV).
On the other hand, there
is CaH absorption at 6382\,\AA, which serves as discriminant 
between dwarfs and giants and is usually seen in 
dwarf stars only (Kirkpatrick et al.~1991, Torres-Dodgen \& Weaver 1993).
We therefore cannot exclude a nature as e.g.~a dwarf nova in 
quiescence.

If the secondary should be a Roche-lobe filling dwarf star of spectral 
type M3, then the distance to RBS1955 would only be $\sim$30\,pc.

Its brightness, unknown nature and the 
likely variability of the source make it a prime 
target for detailed follow-up, with monitoring observations in the 
optical in order to search for possible outbursts, but 
particular in the UV/X-ray range in order to search for a hot 
primary component.

SIMBAD lists RBS1955 as the emission line galaxy UCM 2307+2118 (Zamorano et al.~1996).
The coordinates in the Zamorano paper for the galaxy and for RBS1955 are identical.
The classification of RBS1955 as a CV is undoubted, we guess, that the 
object was misclassified in the earlier publication.

\begin{figure}
\resizebox{\hsize}{!}
{\includegraphics[bbllx=55pt,bblly=85pt,bburx=526pt,bbury=740pt,angle=-90,clip=]{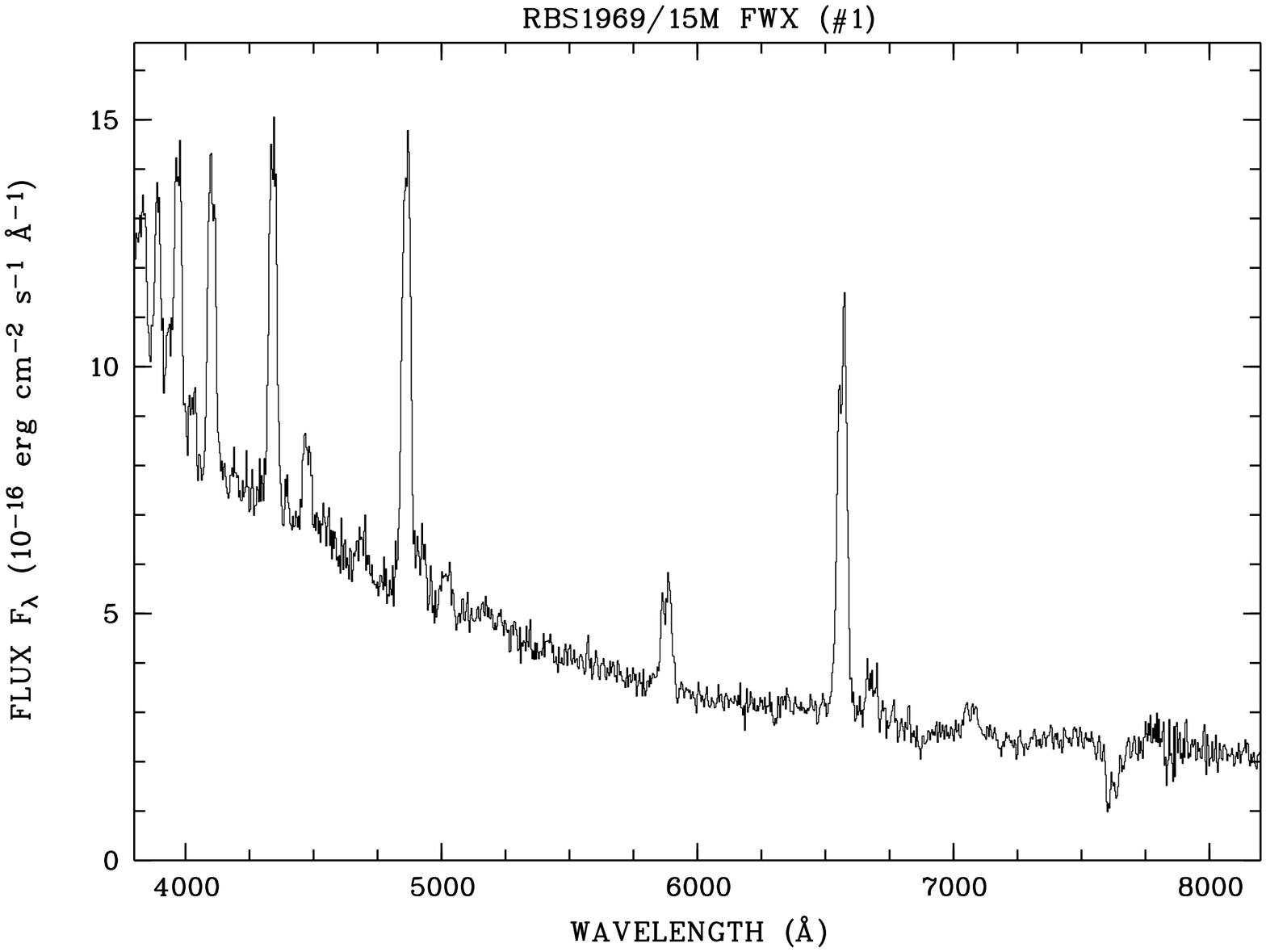}}
\resizebox{\hsize}{!}
{\includegraphics[bbllx=55pt,bblly=85pt,bburx=526pt,bbury=740pt,angle=-90,clip=]{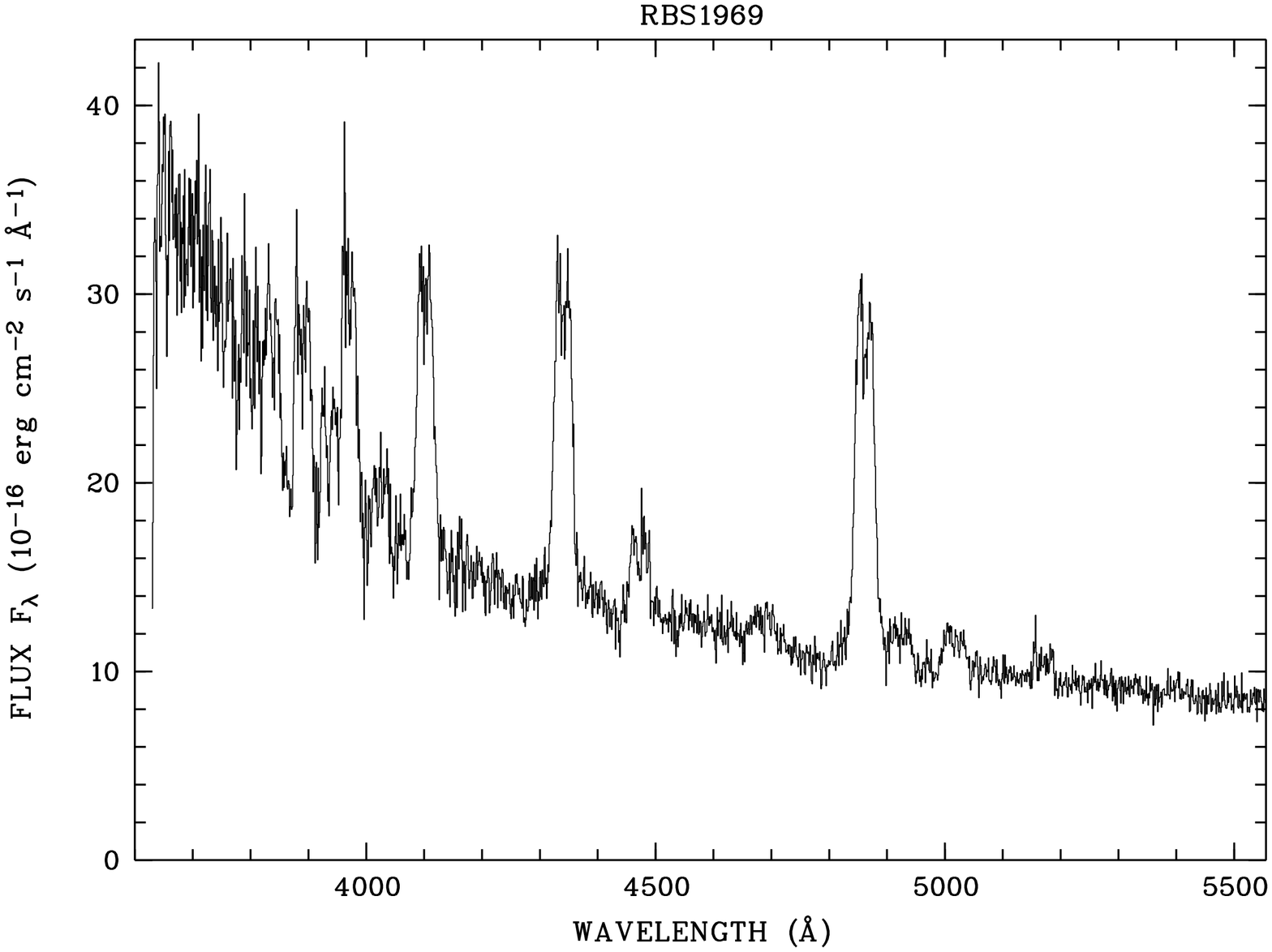}}
\caption{{\it (top)} Identification spectrum of RBS1969, obtained December 2, 1997, 
with the ESO 1.5m telescope. 
{\it (bottom)} Mean blue spectrum of RBS1969, obtained August 23, 1998, 
with the SAAO 1.9m telescope.\label{f:s1969}
}
\end{figure}

\subsection{RBS1969 (= 1RXS\,J231532.3$-$304855)}
Source \#1969 received a very short exposure during the RASS, which allowed to 
collect 21 source photons only. The derived X-ray spectrum is compatible with 
a hard bremsstrahlung component. The discovery spectrum, a 15 min exposure
shown in Fig.~\ref{f:s1969}, was obtained on  November 30, 1997, with the 
ESO 1.5m telescope. Further medium-resolution spectroscopy was obtained from
SAAO on August 23, 1998. All spectra show emission lines
of H-Balmer, HeI and weak HeII together with a Balmer continuum in emission.
The emission lines are double-peaked, which is a typical feature of  a 
highly-inclined accretion disk.
The rather hard X-ray spectrum together with the double-peaked emission lines 
suggest, that RBS1969 is a dwarf nova observed in quiescence at the two 
epochs. Further photometry and spectroscopy are needed in order to determine
the orbital period and to search for a possible eclipse of the disk and the 
white dwarf, which would make the object a prime target for detailed follow-up.

\begin{table*}
\caption{Optical and X-ray data of new non-magnetic CVs}
\label{t:disks}
\begin{tabular}{rccrrccccc}
RBS\# &  EW(H$\beta$)& $M_V$ & $D$  & CRH & $\log F_{0.2-4.0}$ & $\log F_{\rm opt}$ & 
$\log F_{0.2-4.0}/F_{\rm opt}$ &
$(\log F_{0.2-4.0}/F_{\rm opt})^{(c)}_{PR85}$ & $L_{0.2-4.0}$\\
   &  [\AA]          & [mag] & [pc] &[s$^{-1}$]& & & & & [erg s$^{-1}$]\\
\hline
   2 &3$^{(a)}$& 6& 500 & 0.20 &$-11.34$&$-11.16$&$-0.18$&$-1.52$&32.1\\
 372 & 140& 12.3  & 130 & 0.10 &$-11.63$&$-12.55$&$ 0.92$&$ 0.90$&30.7\\
 490 & 230& 13.4  &  33 & 0.18 &$-11.38$&$-11.84$&$ 0.46$&$ 1.21$&29.7\\
1411 &  75& 10.8  & 460 & 0.20 &$-11.34$&$-13.08$&$ 1.74$&$ 0.51$&32.1\\
1955 & --- &--- &30$^{(b)}$&0.13 &$-11.53$& ---     & ---    &    --- &29.5\\
1969 &  70& 10.5  & 165 & 0.18 &$-11.38$&$-12.08$&$ 0.70$&$ 0.47$&31.1\\
\hline
\end{tabular}
\newline
$^{(a)}$ Pattersons relation is invalid for very small EW, 
we used EW = 6 for an estimate of $M_V$\\
$^{(b)}$ distance derived assuming ZAMS secondary\\
$^{(c)}$ estimate based on a relation by Patterson \& Raymond (1985)\\
\end{table*}

\section{On the space density of non-magnetic CVs}
In this section we make an attempt to update the space density of 
cataclysmic variables based on the RBS. We restrict ourselves to the 
non-magnetic systems. Although the magnetic systems are more numerous, 
the RBS catalogue lists 24 polars and 5 intermediate polars, 
they have in general very uncertain distances. Of the 24 polars, about
10 have known or estimated distances, further 4 have lower limits based 
on the non-detection of secondary star features, and further 10 
have no distance estimate at all. A estimate of the 
space density therefore requires more thorough work on the 
distance determination of the individual systems, which is beyond
the scope of this paper.

Following Patterson (1984) we estimate the absolute brightness $M_V$
of the accretion disk in the new non-magnetic CVs from the 
observed equivalent width, EW(H$\beta$), of the H$\beta$ line:
\begin{equation}
\mbox{EW(H}\beta) = 0.3 M_V^2 + e^{0.55(M_V - 4)}.
\end{equation}
Table~\ref{t:disks} lists the relevant numbers for the new CVs.
Although RBS0002 = EC\,23593-6724 was not originally discovered by us
a spectrum was taken in the RBS-program. This is used here to
estimate EW(H$\beta$) and the distance.

\begin{table*}
\caption{Derived quantities of the non-magnetic RBS-CVs. The luminosity in this 
table is the luminosity in the ROSAT spectral band 0.1--2.4\,keV}.
\label{t:nmcvs}
\begin{tabular}{rlccccc}
RBS\# &Name        &  $D$& $\log L_X$& $V_{\rm act}$ & $V_{\rm gen}$ & $V_{\rm act}/V_{\rm gen}$\\
   &               & [pc]&  [s$^{-1}$] & [pc$^{3}$] & [pc$^{3}$] & \\
\hline
   2 &EC 23593-6724& 500:$^{\small r1}$& 32.0& 5.5e+07& 6.3e+07& +0.87 \\
  22 &WW Cet       & 100$^{\small r2}$ & 31.0& 1.5e+06& 6.7e+06& +0.22 \\
 280 &TT ARI       & 135$^{\small r3}$ & 31.1& 3.8e+06& 9.3e+06& +0.40 \\
 288 &WX HYI       & 265$^{\small r4}$ & 31.6& 2.2e+07& 4.1e+07& +0.55 \\
 372 &             & 130$^{\small r1}$ & 30.8& 3.5e+06& 4.3e+06& +0.83 \\
 490 &             &  33$^{\small r1}$ & 29.6& 6.8e+04& 8.3e+04& +0.82 \\
 512 &VW Hyi       &  65$^{\small r4}$ & 30.9& 4.7e+05& 4.7e+06& +0.10 \\
 694 &SU UMa       & 280$^{\small r4}$ & 32.1& 2.0e+07& 8.3e+07& +0.24 \\
 710 &SW UMa       & 140$^{\small r4}$ & 30.9& 3.5e+06& 4.7e+06& +0.76 \\
 713 &EI UMa       &  --               &   --&  --    &  --    &  --\\
 728 &BZ UMa       & 110$^{\small r5}$ & 30.9& 1.9e+06& 4.6e+06& +0.40\\
1008 &T Leo        & 100$^{\small r6}$ & 31.1& 1.6e+06& 7.7e+06& +0.21 \\
1411 &RHS40        & 460$^{\small r1}$ & 32.0& 7.1e+07& 8.8e+07& +0.81 \\
1900 &TY PsA       & 300:$^{\small r1}$& 31.7& 2.3e+07& 3.6e+07& +0.64 \\
1955 &             &  30$^{\small r1}$ & 29.5& 5.3e+04& 5.7e+04& +0.93 \\
1969 &             & 165$^{\small r1}$ & 31.1& 6.5e+06& 1.0e+07& +0.65 \\
\hline
\end{tabular}
\newline
%$^{(a)}$ integrated flux between 0.2 and 4.0 keV in units of 10$^{-12}$\,\ergcmsqs\\
$^{\small r1}$ this work, 
$^{\small r2}$ Verbunt et al.~(1997),
$^{\small r3}$ Patterson (1984),
$^{\small r4}$ Warner (1987),
$^{\small r5}$ Ringwald et al.~(1994),
$^{\small r6}$ Sproats et al.~(1996)
\end{table*}

Assuming that the 
optical emission is dominated by the disk we estimate the distance $d$ 
to the CV from the observed apparent and the derived absolute magnitudes.
We ignore interstellar absorption which is justified by the rather 
small distances.

Fluxes in the X-ray band were derived from the measured RASS count rate 
using PIMMS and a 10 keV thermal bremsstrahlung model with moderate 
absorption of $N_{\rm H} = 5 \times 10^{20}$\,cm$^{-2}$. 
In order to minimize the effect of interstellar absorption, 
one should better use for flux conversion
the count rate in the hard ROSAT band only, CRH $= 0.5 - 2.0$\,keV, 
However, the limiting count rate and limiting flux of our survey 
then would not be properly defined. 
For the analysis of the space density (see below and Table~\ref{t:nmcvs}) 
we therefore stick 
to the count rates in the full ROSAT band. For better 
comparison with the Einstein-IPC we use the count rate in the
hard band CRH.
Conversion factors from count rates to fluxes in the ROSAT and 
Einstein-IPC bands ($0.1-2.4$, $0.2-4.0$\,keV),
are $1.4 \times 10^{-11}$\,\ergcmsqs (PSPC CR cps)$^{-1}$ and 
$2.3 \times 10^{-11}$\,\ergcmsqs (PSPC CRH cps)$^{-1}$, respectively.

For comparison with CVs discovered in the Einstein 
galactic plane survey (Hertz et al.~1990), we list in Table~\ref{t:disks}
the X-ray fluxes in the Einstein band. 
Optical fluxes are derived using $\log F_{\rm opt} = -0.4 V - 5.44$. 
The optical and X-ray data used for the 
flux calculation were not taken simultaneously, hence the flux ratios might 
be incorrect or misleading.
Taken the data in Table~\ref{t:disks} as face values, we can compare the new RBS-CVs
with those found by Hertz et al.~(1990) in the Einstein galactic plane survey (GPX, 
their Fig.~9).
Similar to their new GPX CVs the new RBS-CVs have a 
large H$\beta$ equivalent widths and a high \fxo-ratio. 
Patterson \& Raymond (1985) derived an empirical relation between the 
equivalent widths EW(H$\beta$) and the ratio \fxo, which predicts the 
value of \fxo\ to within a factor of 3. The prediction due to their
relation is listed in the second to last column of Table~\ref{t:disks}.
It agrees reasonably well with the observed values of \#372 and 
\#1969, marginally with \#490, and it disagrees with 
\#0002 and \#1411. Both systems were found to be much brighter 
at X-ray wavelengths (or fainter at optical wavelengths) 
than predicted by the empirical formula. The deviations might
be related to an outburst at the time of the RASS or might 
hint to a somewhat peculiar nature of the sources. 
The luminosities in the IPC-band range from $\log L_{(0.2-4.0)} = 29.5
- 32.1$, i.e.~extending the normal range of CVs to fainter 
luminosities (compare with e.g.~Patterson \& Raymond 1985, their 
Fig.~5).

Similarly to Hertz et al.~(1990) we estimate the space density 
of the non-magnetic CVs in the RBS by a $1/V_{\rm max}$ method.
Since our CVs have high galactic latitude and some of them are at 
a distance in excess of the likely scale height of CVs, 
which we assume to be $h = 200$\,pc, we use the 
modified method by Tinney et al.~(1993). Their method 
of calculation of a generic volume, $V_{\rm gen}$, 
accounts for an exponential density distribution 
$\rho \propto exp[(-d\sin b)/h]$ ($d$: distance, $b$: galactic latitude).
$V_{\rm gen}$ is calculated by 
\begin{equation}
V_{\rm gen} = \Omega \frac{h^3}{\sin^3b}(2 - (\xi^2 + 2\xi +2)e^{-\xi})
\label{e:vgen}
\end{equation}
with $\xi = d \sin b / h$ and $\Omega$ the solid angle of the survey 
($2\pi$ in our case). The maximum generic volume $V_{\rm gen}$ 
is computed using this formula with the 
maximum possible distance of the particular source which 
would allow its detection at the flux limit of the survey.
One particular CV then contributes
$1/V_{\rm gen}$ to the space density $\rho_X$, i.e.~the space density
is $\rho_X = \sum\frac{1}{V_{\rm gen}}$. 
The corresponding numbers for $V_{\rm act}$ and $V_{\rm gen}$ 
are listed in Table~\ref{t:nmcvs}. $V_{\rm act}$ is the volume covered
by this particular source at the measured distance 
according to Eq.~(\ref{e:vgen}).

Using these, the space density of the X-ray selected, non-magnetic,
high-galactic latitude CVs is $\rho_X \ge 3.1 \times 10^{-5}$\,pc$^{-3}$. RBS0713 (= EI UMa)
is missing in the sum due to its unknown distance.
The two new nearby, low-luminosity CVs \#490 and \#1955
have the highest weight in the sum, but both have a somewhat uncertain distance.
Should their distance be much greater than the $\sim$30\,pc derived here,
the lower limit of the space density decreases. 
When omiting these two sources completely, the density becomes 
$\rho_X = 1.5 \times 10^{-6}$\,pc$^{-3}$.

$V_{\rm act}/V_{\rm gen}$ has a uniform distribution between 0 and 1, 
the mean value is  0.56 for the 15 systems with measured or estimated
distances, the uniformity is somewhat higher, if \#490 and \#1955 are
omitted. The mean value then is 0.51.

Our limit of the space density is in good agreement with the recent 
estimate by Patterson (1998), who derives a density of $\sim 10^{-5}$\,pc$^{-3}$.
The reader is refered to his paper, where a thorough discussion of the 
possible selection effects in various surveys is given.

\begin{figure}
\resizebox{\hsize}{!}
{\includegraphics[angle=-90,bbllx=54pt,bblly=63pt,bburx=524pt,bbury=532pt,clip=]{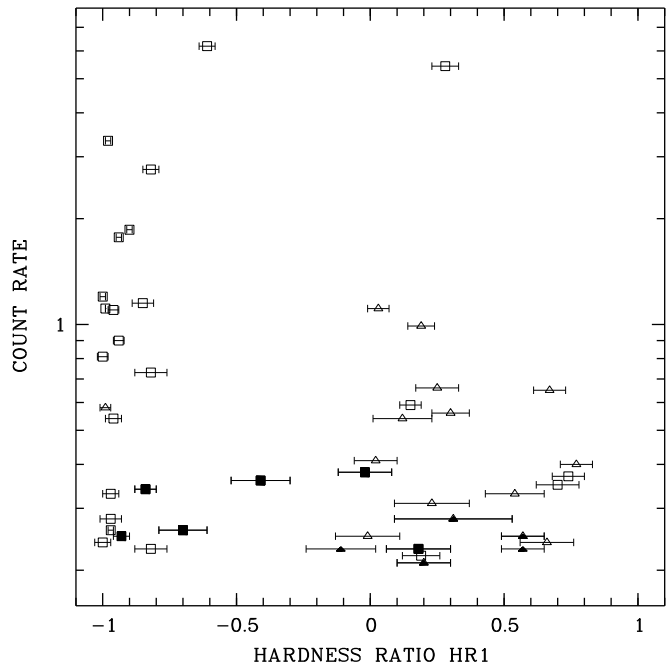}}
\caption{X-ray colour magnitude diagram of all RBS-CVs. Squares indicate magnetic systems
(polars or intermediate polars), triangles indicate non-magnetic systems (dwarf novae, nova-likes
and symbiotics). Filled symbols refer to systems discussed in this paper, open symbols
to the other CVs in the RASS. 
\label{f:rasscmd}
}
\end{figure}

\section{Discussion and conclusions}
We have presented an observational summary of new cataclysmic variables
identified in the RBS. Among the 11 new RBS-CVs are 6 polars, 4 dwarf nova
candiates and one system with uncertain
classification, which was tentatively identified as symbiotic binary but 
could also be a dwarf nova.
Orbital periods could be determined for all polars, they range from 87.1\,min
to 187.7\,min. One of the new polars was found to be asynchronous (RBS1735,
Schwope et al.~1997), another displays a highly peculiar cyclotron spectrum (RBS0206,
Schwope et al.~1999 and this work). Only RBS0206 has a measured field strength,
the other systems do not show cyclotron or Zeeman lines. 

Fig.~\ref{f:rasscmd}, a X-ray colour-magnitude diagram, 
puts the new systems in the context of all CVs in the RBS.
The RBS lists 46 CVs and 1 symbiotic.
Note that these are systems at high-galactic latutide above $|b|>30\degr$ only.
Of the 45 CVs 29 are magnetic (polars
and intermediate polars) and 16 non-magnetic (DN, nova-like).
The Figure shows the known dichotomy between magnetic and non-magnetic systems
which is explained in terms of the typical X-ray spectra of the various 
sub-classes (Beuermann \& Thomas 1993). Polars often have X-ray spectra dominated
by soft X-ray, blackbody-like emission. Three of the new polars (\#206, \#324,
and \#696) follow
the normal pattern, while further three have rather hard X-ray spectra (\#541,
\#1563, \#1735). The reason for this non-conformity is not clear, \#541
and \#1563 appear to be quite normal polars with a one-pole accretion geometry. 
Only \#1735 is unusual, since
it belongs to the rare sub-class of slightly asynchronous polars, where the 
slightly different accretion mode might influence the release of gravitational
energy.

The systems described here enlarge the population at the extremes of 
X-ray/optical flux plane. 
%Almost all CVs described here have an X-ray to 
%optical flux ratio between 1 and 10 
%($F_X = 1\times 10^{-11}$\,erg\,cm$^{-2}$\,s$^{-1}$ per RASS count rate, 
%$\log F_{\rm opt} = -0.4 V - 5.44$). 
Beuermann \& Thomas (1993) plotted all CVs known at that time in the RASS 
count-rate/V-magnitude plane with lines of constant $F_X/F_{\rm opt}$ indicated.
The new magnetic RBS-CVs are located at the low end of the distribution, while the 
new non-magnetic systems are located at the X-ray luminous end of the distribution.
RBS1411 seems to represent the extreme example, it lies
in the domain which is otherwise populated only by the magnetic CVs.

The high $F_{\rm X}/F_{\rm opt}$ ratios of the DN (particularly \#1411) might be
explained by possible outbursts in these systems during the RASS observation.
The spectra of the dwarf nova candidates show no broad absorption
features of the primary star, and these systems are unlikely short period 
CVs of the WZ Sge subclass.

No new obvious IP candidate was found, which
indicates that our census of these intrinsically bright systems is 
already complete at high galactic latitudes. 

We have tentatively identified two new nearby, low-luminosity CVs (\#490 and
\#1955). Whether there is a large population of these sources as predicted 
by Hertz et al.~(1990) is difficult to answer conclusively but our sources
might be the first representatives of this population.

We have discussed the observational properties of the new cataclysmic 
variables found in the RASS. We could determine orbital periods for all
magnetic CVs, period information for the non-magnetic CVs is lacking. 
The prospects of finding 
the periods by photometry is good for \#372, \#490, and
\#1969, since the double-peaked emission lines indicate a moderate to high 
inclination. The two latter systems might show eclipses, the light curves
already obtained for \#372 do not show obvious signs of eclipses.

%We encourage further observations of the optically bright system RBS1955, 
%which could not be uniquely identified. RBS1411 has an extreme X-ray to 
%optical flux ratio and is a promising target for future X-ray observations.

\acknowledgements
We acknowledge helpful comments by Dr.~U.~Munari (Padova) and N. Lodieu 
(AIP) on the spectrum of RBS1955. We thank our referee, Dr.~D.~Christian,
for his comments, which helped to improve the paper.
This project was supported by the DLR through grant 50 OR 9706 8.

The ROSAT project is supported by the Bundesministerium f\"{u}r 
Bildung, Wissenschaft, Forschung und Technologie (BMBF/DLR) and the 
Max-Planck-Gesellschaft. We thank the ROSAT team for performing the All-Sky
Survey and producing the RASS Bright Source Catalogue. 

This research has made use of the SIMBAD database operated at
CDS, Strasbourg, France, and the NASA/IPAC Extragalactic database (NED)
operated by the Jet Propulsion Laboratory, California Institute of 
Technology under contract with the National Aeronautics and Space 
Administration. Identification of the RASS X-ray sources was greatly
facilitated by use of the finding charts based upon the COSMOS
scans of the ESO/SERC J plates performed at the Royal Observatory 
Edinburgh and APM catalogue based on scans of the red and blue POSS plates
performed at the Institute of Astronomy, Cambridge, UK.

Based in part 
on photographic data of the National Geographic Society -- Palomar
Observatory Sky Survey (NGS-POSS) obtained using the Oschin Telescope on
Palomar Mountain.  The NGS-POSS was funded by a grant from the National
Geographic Society to the California Institute of Technology.  The
plates were processed into the present compressed digital form with
their permission.  The Digitized Sky Survey was produced at the Space
Telescope Science Institute under US Government grant NAG W-2166.

\end{document}